\documentclass[pra,twocolumn,superscriptaddress,nobalancelastpage]{revtex4}
\usepackage{graphicx}
\usepackage{color}
\usepackage{amsmath}
\usepackage{color}
\usepackage{float}

\allowdisplaybreaks[1]

\begin{document}

\title{Filtering of the absolute value of photon-number difference\\
  for two-mode macroscopic quantum superpositions}

\author{M. Stobi\'nska}
\affiliation{Institute of Theoretical Physics and Astrophysics, University of Gda\'nsk, ul. Wita Stwosza 57, 80-952 Gda\'nsk, Poland}
\affiliation{Institute of Physics, Polish Academy of Sciences, Al.\ Lotnik\'ow 32/46, 02-668 Warsaw, Poland}

\author{F. T\"oppel}
\affiliation{Max Planck Institute for the Science of Light, Guenther-Scharowsky-Str. 1/Bldg. 24, 91058 Erlangen, Germany}
\affiliation{Institute for Optics, Information and Photonics, University of Erlangen-N\"urnberg, Staudtstr. 7/B2, 91058 Erlangen, Germany}

\author{P. Sekatski}
\affiliation{Group of Applied Physics, University of Geneva, Chemin de Pinchat 22, CH-1211 Geneva, Switzerland}

\author{A. Buraczewski}
\affiliation{Faculty of Electronics and Information Technology, Warsaw University of Technology, ul. Nowowiejska 15/19, 00-665 Warsaw, Poland}

\author{M.~\.Zukowski}
\affiliation{Institute of Theoretical Physics and Astrophysics, University of Gda\'nsk, ul. Wita Stwosza 57, 80-952 Gda\'nsk, Poland}
\affiliation{University of Science and Technology of China, Hefei, Anhui, China}

\author{M. V. Chekhova}
\affiliation{Max Planck Institute for the Science of Light, Guenther-Scharowsky-Str. 1/Bldg. 24, 91058 Erlangen, Germany}
\affiliation{Department of Physics, M.~V.~Lomonosov Moscow State University, Leninskie Gory, 119991 Moscow, Russia}

\author{G. Leuchs}
\affiliation{Max Planck Institute for the Science of Light, Guenther-Scharowsky-Str. 1/Bldg. 24, 91058 Erlangen, Germany}
\affiliation{Institute for Optics, Information and Photonics, University of Erlangen-N\"urnberg, Staudtstr. 7/B2, 91058 Erlangen, Germany}

\author{N. Gisin}
\affiliation{Group of Applied Physics, University of Geneva, Chemin de Pinchat 22, CH-1211 Geneva, Switzerland}
  
\date{\today}
\pacs{42.50.Dv,42.65.Yj,03.67.-a}

\begin{abstract}
We discuss a device capable of filtering out two-mode states of light with mode populations differing by more than a certain threshold, while not revealing which mode is more populated. It would allow engineering of macroscopic quantum states of light in a way which is preserving specific superpositions. As a result, it would enhance optical phase estimation with these states as well as distinguishability of ``macroscopic'' qubits. We propose an optical scheme, which is a relatively simple, albeit non-ideal, operational implementation of such a filter. It uses tapping of the original polarization two-mode field, with a polarization neutral beam splitter of low reflectivity. Next, the reflected beams are suitably interfered on a polarizing beam splitter. It is oriented such that it selects unbiased polarization modes with respect to the original ones. The more an incoming two-mode Fock state is unequally populated, the more the polarizing beam splitter output modes are equally populated. This effect is especially pronounced for highly populated states. Additionally, for such states we expect strong population correlations between the original fields and the tapped one. Thus, after a photon-number measurement of the polarizing beam splitter outputs, a feed-forward loop can be used to let through a shutter the field, which was transmitted by the tapping beam splitter. This happens only if the counts at the outputs are roughly equal. In such a case, the transmitted field differs strongly in occupation number of the two modes, while information on which mode is more populated is non-existent (a necessary condition for preserving superpositions).  
\end{abstract}

\maketitle

\section{Introduction}

The set of efficiently produced quantum states of light is limited. It is especially difficult to produce non-classical non-Gaussian superpositions. Nevertheless, with quantum state engineering certain properties of accessible states can be modified or enhanced.  In particular, measurement induced state operations which facilitate preparing a quantum state for some further tasks, allow filtering out states of required features and may lead to non-Gaussian characteristics of the resulting states. Often, they involve intensity measurements, for which crucial are threshold detectors, selecting Fock states or their superpositions with sufficiently high population.  Examples of low-threshold detectors are realized with single photon on-off detectors or human eyes~\cite{Sekatski2009,Sekatski2010}. They can be applied in setups that perform POVM measurements~\cite{POVM} leading to quantum operations. As a result, it is possible to block light of unwanted properties (too low or too high intensity). More complicated filters for Fock states utilize interference effects~\cite{Sanaka2006,Resch2007}. A more challenging task is to construct a filter selecting states of certain properties (on request), while preserving quantum superpositions. This is very important for superpositions of the Schr\"odinger-cat type. 

Recently, macroscopic quantum superpositions became experimentally accessible for light in the form of the micro-macro singlet state~\cite{DeMartini2008} and the entangled bright squeezed vacuum~\cite{Macrobell}.  In the former state, produced by optimal quantum cloning, a single photon is entangled with a ``macroscopic'' qubit in a polarization singlet state. The latter is a macroscopic analog of two-photon polarization Bell states~\cite{Masha2}. Since these states combine quantum properties with macroscopic population and could enable efficient light-matter coupling, they are interesting for quantum information technology: quantum memory~\cite{Appel2008,Burks2009,Gerasimov2011}, quantum key distribution~\cite{Gisin2002}, quantum metrology~\cite{Vitelli2010-3,SpagnoloNoisy} and macroscopic Bell tests~\cite{Vitelli2010-2,Sekatski2011}. However, their distinguishability is low in analog detection and they are easily destroyed by losses~\cite{Stobinska09,Stobinska2011,Buraczewski2011,Vitelli2010}. Special quantum state filtering applied to these states gives hope to solve the problem of detection and to enhance their properties useful for quantum technology tasks.

We present a theory of a device capable of filtering out two-mode states of light with mode populations differing by more than a certain threshold.
We call it modulus of intensity difference filter (MDF). It performs a non-Gaussian operation and works as quantum scissors~\cite{scissors} for general two-mode Fock state superpositions. We show that, effectively, MDF filters out superpositions of N00N-like components, allowing an enhanced optical phase estimation with macroscopic quantum states of light. We also show that it improves distinguishability of ``macroscopic'' qubits in realistic scenarios.

We propose a simple optical scheme, which gives an approximate operational implementation of such a filter for two orthogonal (linear) polarization modes. The field is fed into a polarization neutral (tapping) beam splitter of low reflectivity. The weak reflected modes are suitably interfered on a polarizing beam splitter oriented such that it selects diagonal and anti-diagonal polarization modes with respect to the original ones. The more an incoming  two-mode Fock state is unequally populated, the more the output modes are roughly equally populated. Since the reflected and transmitted beams are correlated, estimating the modulus of population difference for the former gives an estimate for the latter. This effect is especially pronounced for highly populated states. After a photon-number measurement of the outputs of the polarizing beam splitter, a feed-forward loop can be used to let through a shutter the field, which was transmitted by the tapping beam splitter, only in the case of roughly equal counts at the outputs. Such a field differs strongly in occupation number of the two modes, while information on which mode is more populated is non-existent. Thus, a necessary condition for preserving superpositions is satisfied.

The paper is organized as follows. In Section \ref{sec2} we discuss the theoretical description and properties of modulus of intensity difference filter. In Section \ref{sec3} we analyze the action of the theoretical MDF on ``macroscopic'' qubits, a part of micro-macro polarization singlet state. Section \ref{sec4} is devoted to the operational scheme giving effectively  an MDF.

\section{Theory and properties of MDF}\label{sec2}     

We define an  MDF  as a device which performs the following  projection operation
\begin{equation}
  \mathcal{P}_{\delta_{th}} = \sum_{\substack{k,l=0;\lvert k-l\rvert\ge {\delta}_{th}}}^{\infty} |k,l\rangle \langle k,l|,
  \label{theoretical-preselection}
\end{equation}
where $|k,l\rangle$ is a two-mode Fock state. For simplicity, let us consider polarization modes. If ${\delta}_{{th}}>0$ the filter acts as ``quantum scissors''~\cite{scissors}. It cuts out those Fock components for which the modulus of occupation difference is below the threshold ($|k-l|< \delta_{th}$), and preserves the ones with the modulus of difference above it ($|k-l| \ge \delta_{th}$).

\begin{figure}
  \begin{center}
    \raisebox{2cm}{(a)}
    \includegraphics[height=3cm]{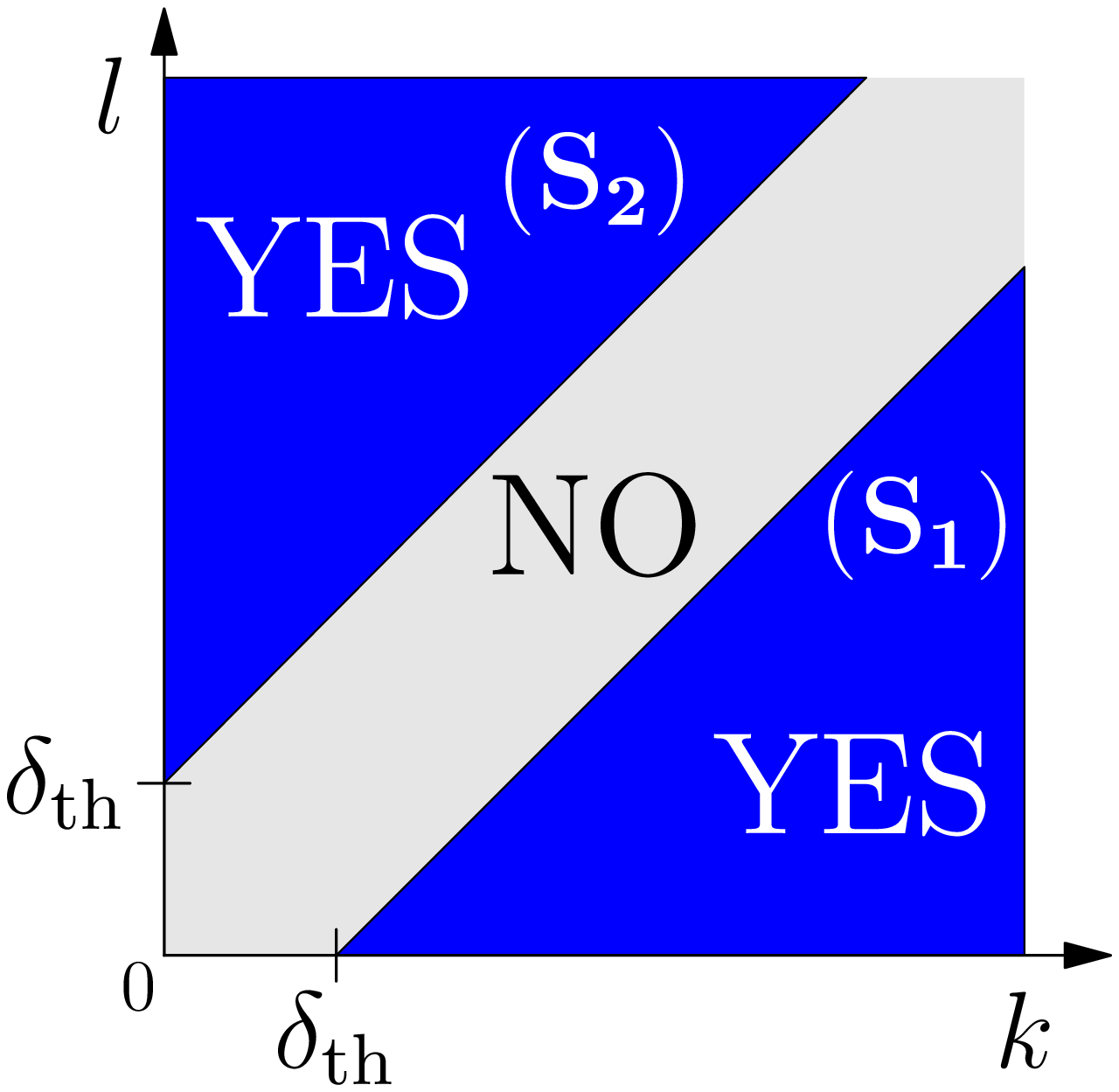}\hskip0.5cm
    \includegraphics[height=3cm]{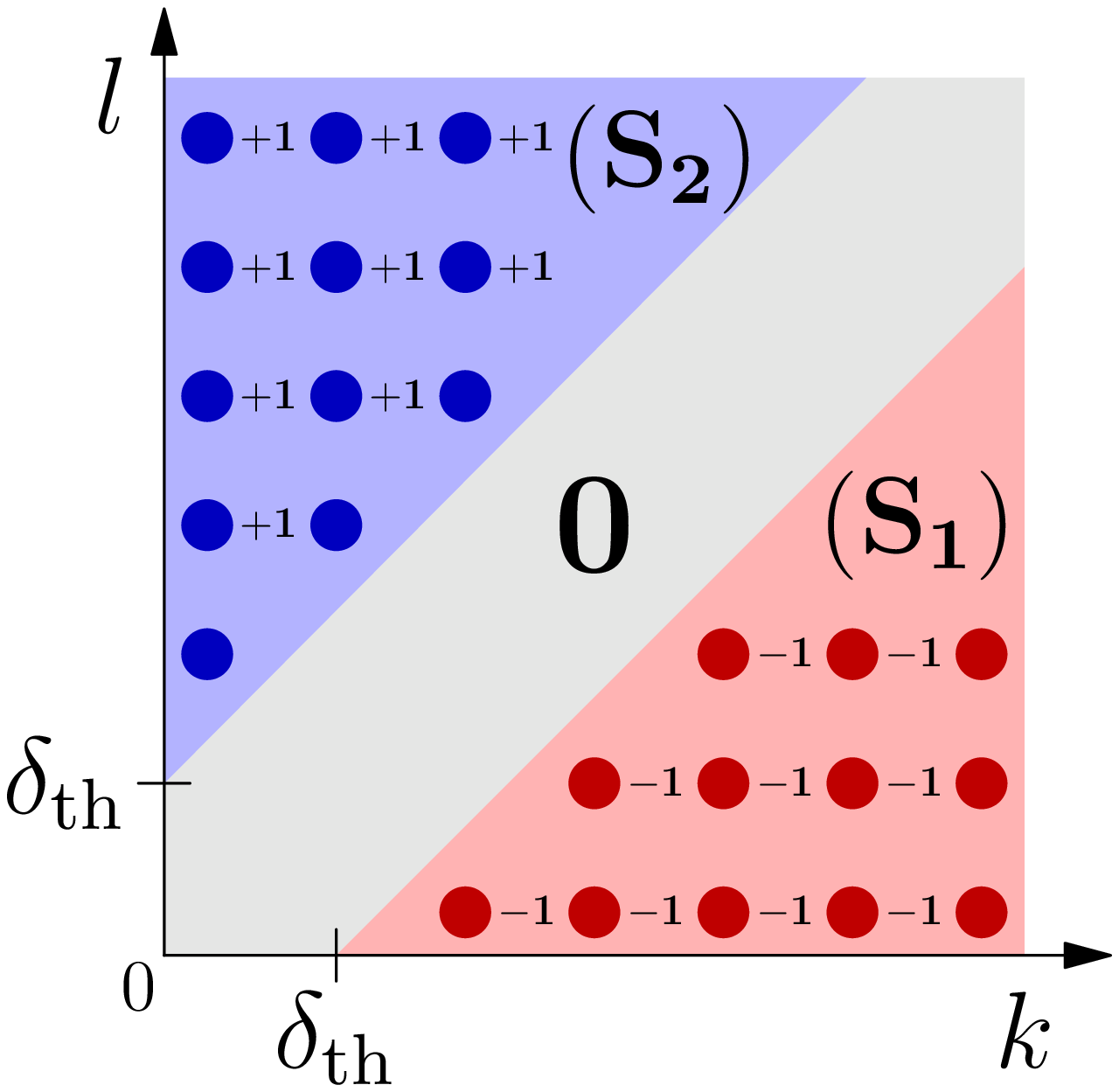}
    \raisebox{2cm}{(b)}\\
   \end{center}
  \caption{Comparison of two filtering techniques: absolute difference (MDF) (a) and orthogonality filter (OF) (b). The dots in (b) symbolize specific possible measurement results of photon numbers. The state of the field filtered by an OF is represented by one of the dots. In the case of MDF, the state is projected onto the whole YES region, which preserves quantum coherence of components occupying both regions. $k$ and $l$ denote numbers of photons in two orthogonal polarization modes.}
  \label{fig:filtering}
\end{figure}

We would like to comment on two key features of the filter. First of all, it estimates the absolute value of the difference instead of the difference. This procedure is experimentally more demanding, but it has an advantage. Since all non-zero eigenvalues of the operator $\mathcal{P}_{\delta_{th}}$ are equal to $1$, the filter does not provide any information on which polarization mode was more populated. Thus, if a qubit is encoded in highly populated polarization states, like e.g. in Eq.~(\ref{macro-qubits}), it does not discriminate these states and filters them fairly. This property is important for all quantum protocols requiring state preparation without the state readout. The other main feature is that the filtering is performed in a ``yes''-``no'' manner: the exact value of the modulus is {\it never} measured. This is a key property for quantum protocols which require engineering preserving the superposition. For these reasons we call this device a filter.

These features are the main difference between the MDF and the orthogonality filter (OF) executing direct intensity difference measurements~\cite{DeMartini-PRL}. The OF is the basic element in setups performing measurement induced operations on macroscopic polarization states~\cite{Vitelli2010-2}.  Contrary to the MDF which performs a non-destructive measurement, the OF destroys superpositions and allows only for efficient state discrimination in detection, not filtering, and is not suitable for preselection strategies in Bell tests~\cite{Vitelli2010-2}. In the case of a micro-macro singlet, it identifies the state and breaks entanglement. The action of the MDF and OF is compared in Fig.~\ref{fig:filtering}. MDF projects onto $S_1$ and $S_2$ area.  Superpositions  of components belonging to $S_1$ and $S_2$ are preserved. OF, combined with photomultipliers, projects the state on a Fock state either in $S_1$ or $S_2$, illustrated as a red or blue dot in Fig.~\ref{fig:filtering}.

\section{Filtering of ``macroscopic'' qubits}\label{sec3}

Let us analyze the action of the operator $\mathcal{P}_{\delta_{th}}$ on specific ``macroscopic'' qubits (macro-qubits), which are the macroscopic part of micro-macro polarization singlets. They are produced by optimal phase covariant quantum cloning via phase sensitive parametric amplification~\cite{Sekatski2010,DeMartini-PRL,Masha} of single photons of a defined polarization ($\varphi$ or $\varphi^\perp$, respectively)
\begin{eqnarray}
|\Phi\rangle &=& \sum_{i,j=0}^{\infty} \!\gamma_{ij}
\big|2i+1,2j\rangle,
\label{macro-qubits}\\
|\Phi_{\perp}\rangle &=& \sum_{i,j=0}^{\infty} \!\gamma_{ij}
\big|2j,2i+1\rangle, \nonumber \label{MACROQUBIT}
\end{eqnarray}
where e.g. states $|k,l\rangle$ represent $k$ photons in polarization state $|\varphi\rangle$, and $l$ in $|\varphi^{\perp}\rangle$, which in turn are defined as $|\varphi\rangle=(e^{i\varphi}|H\rangle+ e^{-i\varphi}|V\rangle)/\sqrt{2}$ and $|{\varphi^\perp}\rangle=i(e^{i\varphi}|H\rangle-e^{-i\varphi}|V\rangle)/\sqrt{2}$~\cite{Sekatski2010}, where $H$ and $V$ represent linear horizontal and vertical polarizations. The probability amplitudes equal  $\gamma_{ij}=\cosh g^{-2} \left((\tanh g)/2\right)^{i+j} \sqrt{(1+2i)!(2j)!}/i!/j!$, where $g$ is the parametric gain. Due to a different parity of  occupation numbers of the two polarizations, the states $|\Phi\rangle$ and $|\Phi_{\perp}\rangle$ are orthogonal. 

In a recent experiment~\cite{DeMartini-PRL}, realizations of such states contained up to $4\sinh^2 g \simeq 10^{4}$ photons on average. However, in high photon number regime the detectors are not single photon resolving, but distinguish counts varying by at least $\pm 150$ photons ~\cite{Masha}. Thus, macro-qubits are hardly distinguishable with direct detection~\cite{DeMartini-PRL}. 

To overcome this problem, an MDF could be used to enhance the distinguishability. Two important traits of the states are crucial. The average number of photons in polarization $\varphi$ in $|\Phi\rangle$ is three times higher than the number of photons in polarization $\varphi^{\perp}$, and {\em vice versa} for $|\Phi_{\perp}\rangle$. Further, if one excludes superposition components with approximately identical numbers of photons in the two polarizations, this ratio increases. Thus, an MDF would definitely increase the distinguishability of the states.

Imagine a scheme which uses an MDF, and behind it we place detection station which measures number of photons in the two polarization modes. In such a case the distinguishability may be quantified in terms of photon distributions $p_{\Phi}(k,l)=|\langle k,l|\Phi\rangle|^{2}$ and $p_{\Phi_{\perp}}(k,l)$ giving the probabilities of finding simultaneously $k$ photons in polarization $\varphi$ and $l$ in~$\varphi^{\perp}$. For the filtered macro-qubits with the operator $\mathcal{P}_{\delta_{th}}$ they equal (see Appendix~A) 
\begin{equation}
p_{\Phi}(k,l) \!=\kern-2em\sum_{\substack{i,j=0;\lvert 2i+1-2j \rvert\ge {\delta}_{th}}}^{\infty} \kern-2em{\tilde{\gamma}}_{ij}^2\, \delta_{k, 2i+1} \delta_{l, 2j}, \;
p_{\Phi_\perp}(k,l) \!=\! p_{\Phi}(l,k),
\end{equation}
where $\tilde{\gamma}_{ij}$ are renormalized $\gamma_{ij}$, and $\delta_{a,b}$ is the Kronecker delta. Since the distribution $p_{\Phi_\perp}$ is mirror reflected with respect to $p_{\Phi}$ along the $k=l$ line, we divide the space $(k,l)$ into two triangular areas $S_1$ for $k\ge l$ and $S_2$ for $k<l$. The distinguishability reads
\begin{equation}
\label{eq:distinguishability}
v = P_{\Phi}^{(S_1)}-P_{\Phi_{\perp}}^{(S_1)} = P_{\Phi}^{(S_1)}-P_{\Phi}^{(S_2)},
\end{equation}
where $P_{\Phi}^{(S_i)}= \sum_{k,l \in S_i}p_{\Phi}(k,l)$ is the probability of finding $|\Phi\rangle$ in $S_i$ and $P_{\Phi}^{(S_1)}+P_{\Phi}^{(S_2)}=1$. It increases if $|\Phi\rangle$ ($|\Phi_{\perp}\rangle$) starts to occupy mostly one of $S_i$ regions, e.g.\ $S_1$ ($S_2$), with increasing $\delta_{th}$. Fully distinguishable (indistinguishable) states have $v=1$ ($v=0$).

\begin{figure}
  \begin{center}
    \raisebox{4cm}{(a)}
    \includegraphics[height=4.5cm]{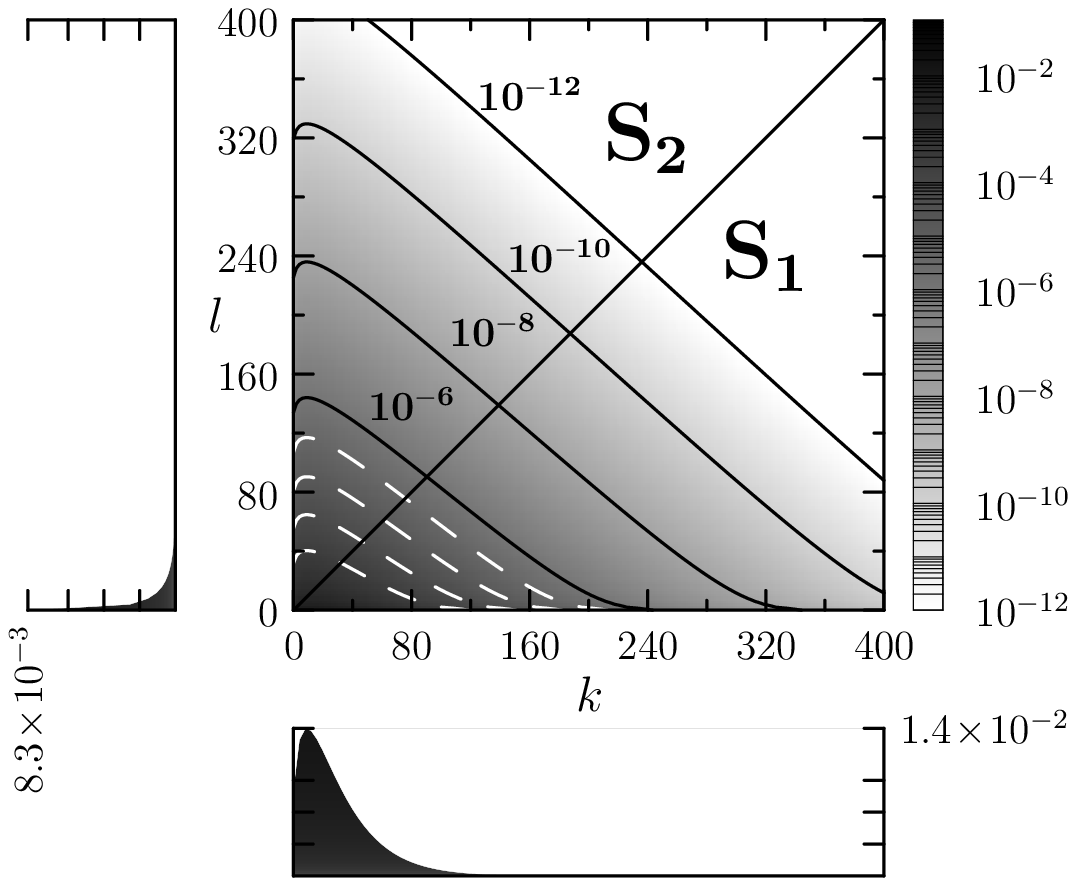}\\[5mm]
    \raisebox{4cm}{(b)}
    \includegraphics[height=4.5cm]{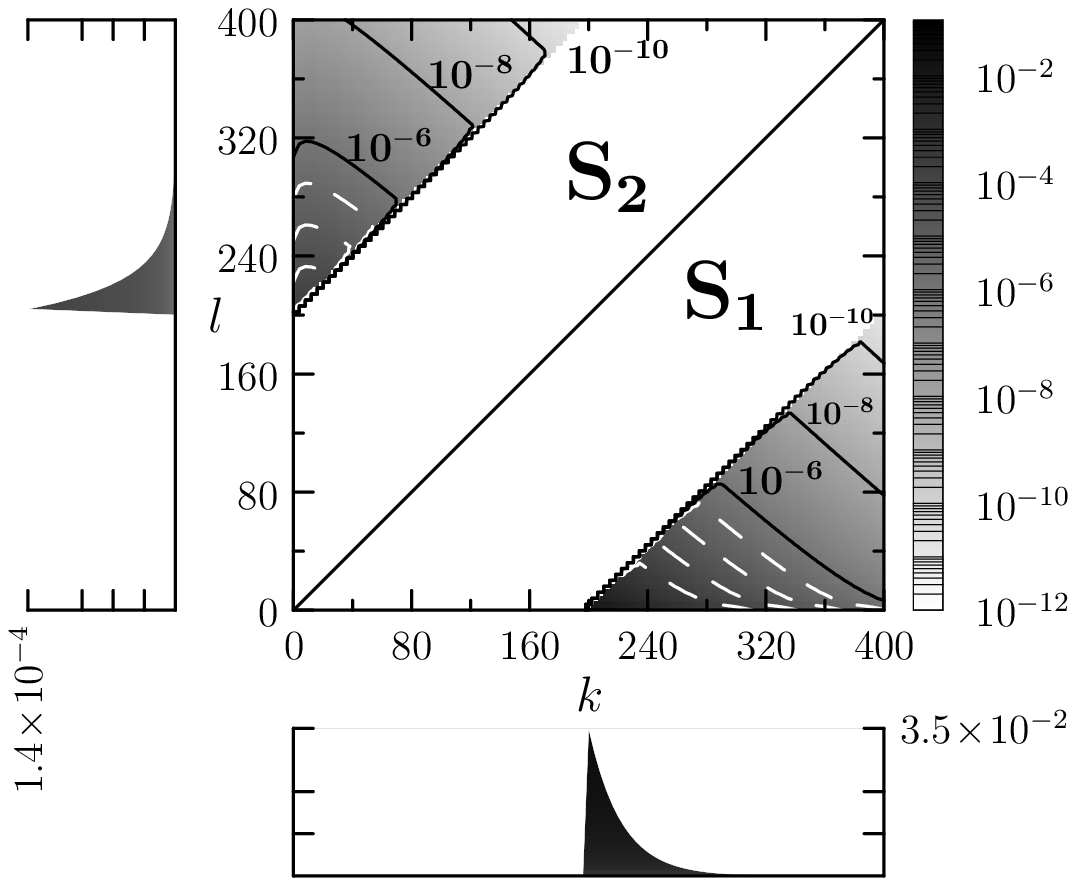}\\
  \end{center}
  \caption{Photon distribution $p_{\Phi}$ for the macroscopic state $|\Phi\rangle$ computed for $g=1.87$ and filtering threshold $\delta_{th}=0$ (a) and $\delta_{th}=200$ (b). $k$ and $l$ denote numbers of photons in two orthogonal polarization modes. The one-dimensional plots show values of $p_{\Phi}$ for $k=0$ (the left one) and $l=0$ (the bottom one), respectively.}
  \label{fig:qfuncIdeal}
\end{figure}

Originally, the photon-number distribution $p_{\Phi}(k,l)$ occupies both $S_1$ and $S_2$ and is almost equally distributed between them giving $v=0.64$, independently of the gain $g$, see Fig.~\ref{fig:qfuncIdeal}a. Fig.~\ref{fig:qfuncIdeal} is plotted for $g=1.87$. The filtering cuts out a stripe, $\sqrt{2} \delta_{th}$ wide, located symmetrically along the $k=l$ line. In Fig.~\ref{fig:qfuncIdeal}b we took $\delta_{th}=200$.  The state $|{\Phi}\rangle$ occupies two disjoint regions of space: the bottom ($S_1$) and top ($S_2$) triangles, but increasing the threshold from $ \delta_{th}=0$ to $ \delta_{th}=200$ reduces the contribution of $p_{{\Phi}}$ in $S_2$: the peak value goes down originally from $8.3 \cdot 10^{-3}$ to $1.4 \cdot 10^{-4}$. Simultaneously, the distribution peak in $S_1$ increases from $1.4 \cdot 10^{-2}$ to $3.5 \cdot 10^{-2}$.  Similar behavior is observed for higher gains. The behavior of $p_{{\Phi}_\perp}$ is identical but mirror reflected. Thus, distinguishability increases.

\begin{figure}
  \begin{center}
    \raisebox{4cm}{(a)}
    \includegraphics[height=4.5cm]{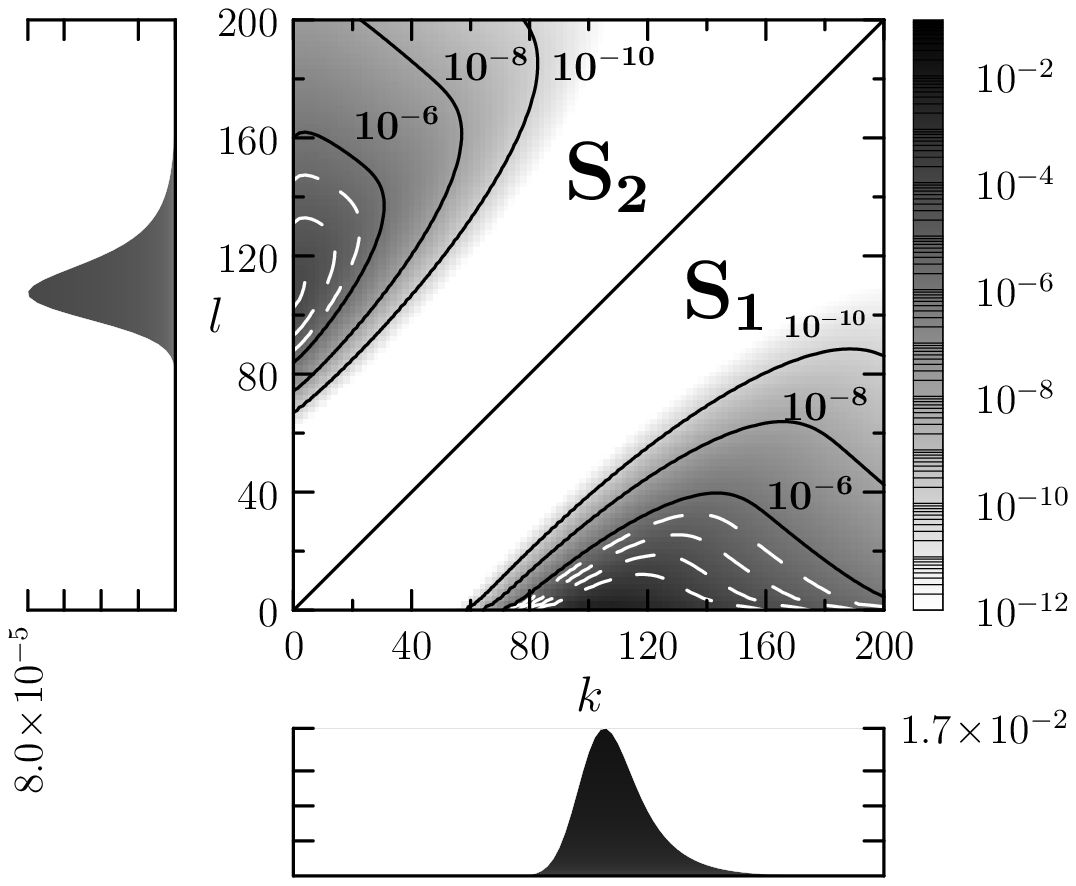}\\[5mm]
    \raisebox{4cm}{(b)}
    \includegraphics[height=4.5cm]{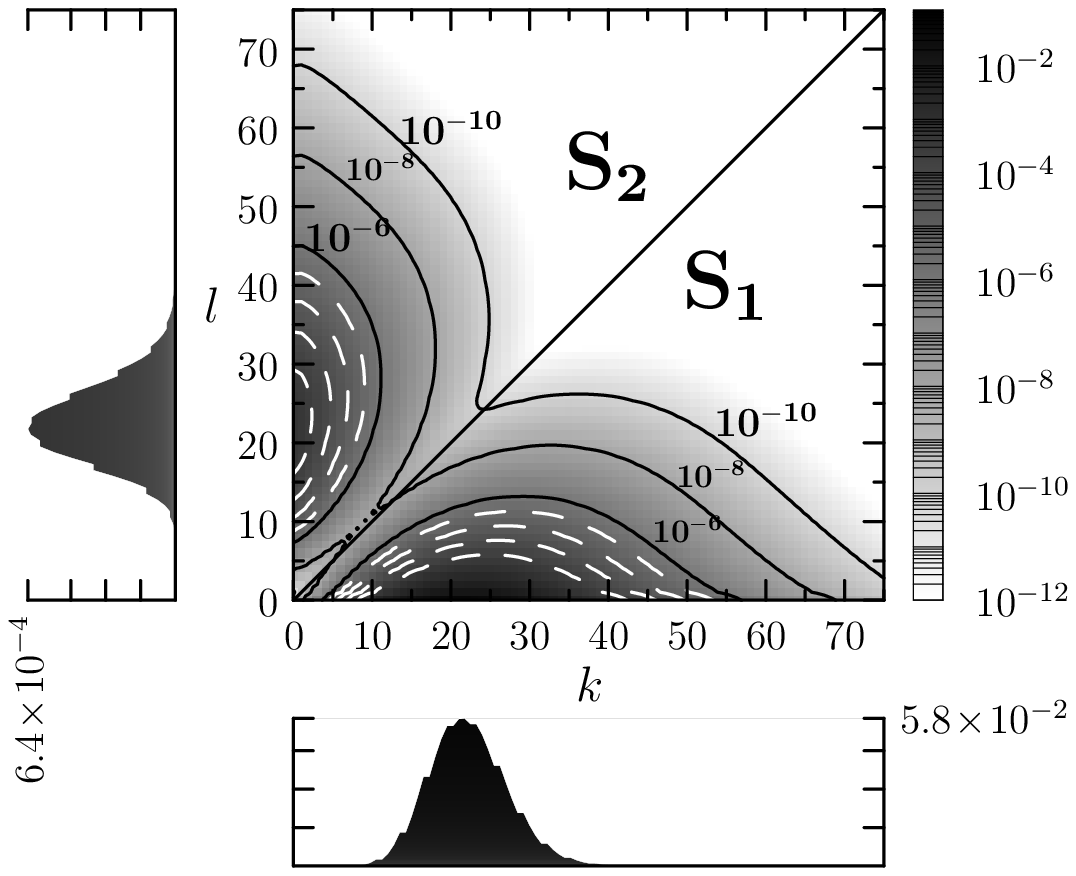}
  \end{center}
  \caption{Photon distribution $p_{\Phi}$ for the macroscopic state $|\Phi\rangle$ computed for $g=1.87$, filtering threshold $\delta_{th}=200$ and $50\%$ (a) and $90\%$ (b) of losses. $k$ and $l$ denote numbers of photons in two orthogonal polarization modes. The one-dimensional plots show values of $p_{\Phi}$ for $k=0$ (the left one) and $l=0$ (the bottom one), respectively.}
  \label{fig:qfunc}
\end{figure}
The effect of increased distinguishability remains even in the presence of losses. The losses can be modeled by a beam splitter (BS) with a reflectivity $R$ (see Appendix A) put in front of an ideal detector. The $p_{{\Phi}}$ distributions evaluated for $g=1.87$, $\delta_{th}=200$ and $50\%$ and $90\%$ of losses are depicted in Fig.~\ref{fig:qfunc}. The loss results in shifting the distribution towards the origin of the coordinates, i.e.\ the vacuum state. The distribution peaks become smooth and symmetric. The edges along the threshold lines are blurred and the bigger the losses, the smaller the width of the gap. It disappears completely for $90\%$ of losses. With increasing losses the height of the upper and left peak first drops, and next increases, because the total probability over the whole space $(k,l)$ has to be 1.

\begin{figure}
  \begin{center}
    \includegraphics[height=5cm]{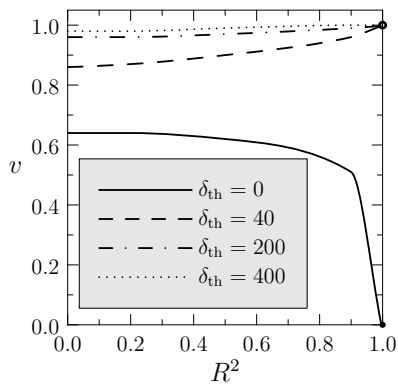}\\
   \end{center}
   \caption{Distinguishability $v$ of macro-qubits (Eq.~(\ref{eq:distinguishability})) evaluated for gain $g=1.87$ and several threshold values $\delta_{{th}}$ as function of losses $R$.}
  \label{fig:visibilities}
\end{figure}
For states (\ref{macro-qubits}) we have numerically computed their distinguishability $v$ for gain $g=1.87$ and several filtering thresholds $\delta_{{th}}$ as a function of losses, see Fig.~\ref{fig:visibilities}. If no filtering is applied, then $v=0.64$, but drops quickly to $0$ if $R>0.9$. If $\delta_{th}$ increases, $v$ increases as well and approaches unity with a reasonable probability of success, e.g.\ $v=0.96$ with $p_s=10^{-4}$. Obviously, for $R=1$ the states become vacuum and we get $v=0$ independently of $\delta_{th}$ (this is indicated by an open circle in the upper curves and a full circle in the solid line in Fig.~\ref{fig:visibilities}).

\section{Simple operational scheme for approximate MDF} \label{sec4}

\begin{figure}
\begin{center}
  \includegraphics[width=5cm]{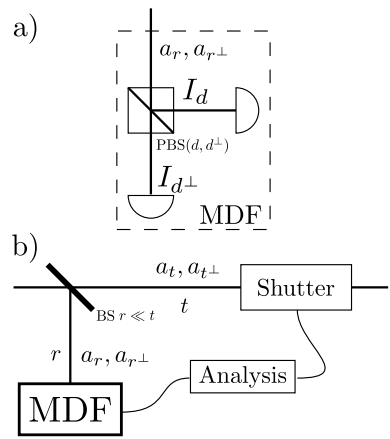}
\end{center}
\caption{An approximate operational scheme of an MDF. The box MDF in (b) is the setup given in (a). The details are in the main text.}
\label{fig:MZI}
\end{figure}

Our scheme for an approximate realization of an MDF for polarization modes is shown in Fig.~\ref{fig:MZI}a. The setup in Fig.~\ref{fig:MZI}b shows its application for the measurement induced operations on quantum states. It uses tapping of the original field, with a polarization neutral BS of a low reflectivity (Fig.~\ref{fig:MZI}b). The reflected beams, $a_r$, $a_{r^{\perp}}$ are suitably interfered on a polarizing beam splitter (PBS) oriented such that it selects unbiased polarization modes with respect to the original ones (Fig.~\ref{fig:MZI}a). The more an incoming  two-mode Fock state is unequally populated, the more the output modes are roughly equally populated. This effect is especially pronounced for highly populated states, and additionally for such states we expect strong population correlations between the original fields and the tapped one. Thus, after a photon-number measurement of PBS outputs, a feed-forward loop can be used to let through a shutter the field, that was transmitted by the tapping BS. This happens only in the case of roughly equal counts at the outputs. Such a field differs strongly in occupation number of the two modes, while information on which mode is more populated is non-existent (a necessary condition for preserving superpositions).  

Let us move to the details of operation of the part of the device shown in Fig.~\ref{fig:MZI}a. A two-mode $r$, $r^{\perp}$ polarization light beam enters PBS which works in a basis $d$, $d^{\perp}$ unbiased with respect to the basis in which we write the original superposition. For example, the beam could be defined in diagonal/ anti-diagonal basis, while PBS may select in left-handed/right-handed polarization basis. Let us denote the annihilation operators of the polarization modes entering PBS by $a_{r}$, $a_{r^{\perp}}$. PBS transforms them according to the unitary operation such that its output mode operators equal $a_d=1/\sqrt{2}(a_r+a_{r^{\perp}})$, $a_{d^{\perp}}=1/\sqrt{2}(a_{r^{\perp}}-a_r)$. The two orthogonally polarized exit beams $d$ and $d^{\perp}$, propagate to a pair of detectors, which measure their photon numbers $I_{d}=K$ and $I_{d^{\perp}}=L$.

We will examine the work of the setup (Fig.~\ref{fig:MZI}a) by its action on a general two-mode polarization input state which is a Fock state 
%$\lvert\Psi_{in}\rangle =\sum_{n,m}\xi_{n m}\lvert n,m\rangle_r$. 
$\lvert n,m\rangle_r$. 
Detection behind PBS projects this state onto a two-mode Fock state $|K,L\rangle_d = \tfrac{1}{\sqrt{K! L!}} {a^{\dagger}_d}^K {a^{\dagger}_{d^{\perp}}}^L |0\rangle$. The states $|n,m\rangle_r$ form a basis in the considered subspace of photon states. Note, that one can introduce a different indexation of the basis, namely $|\tfrac{1}{2}(S_r+\Delta_r),\tfrac{1}{2}(S_r-\Delta_r)\rangle_r$, where $S_r=n+m$ and $\Delta_r=n-m$, which is one-to-one. Let us denote such basis states $|\Psi^{S_r,\Delta_r}\rangle_r$. The states $|K,L\rangle_d$ also form such a basis, which is related to the previous one via the unitary transformation of BS. The probability of obtaining $|K,L\rangle_d$ from $|\Psi^{S_r,\Delta_r}\rangle_r$ input is $p(K,L|S_r,\Delta_r)=|\langle \Psi^{S_r,\Delta_r}|K,L\rangle_d|^2$. However, $p(K,L|S_r,\Delta_r)=p(S_r,\Delta_r|K,L)$ due to the bi-stochastic nature of such quantum probabilities~\cite{Peres}. Note that the measured total number of photons $S = K+L$, if the initial state is $|\Psi^{S_r,\Delta_r}\rangle_r$, must be $S=S_r$. Let us change the variables $L$ and $K$, so that they would correspond to the quantities useful for the further analysis of the filtering: the total sum $S$ and the population difference $\Delta=L-K$ of the registered photons. The probability distribution of the occupation difference $\Delta_r$ in the incoming modes $r$ and $r^{\perp}$ given that $S$ and $\Delta$ were measured $p^{S,\Delta}(\Delta_r) = p(S_r, \Delta_r|\tfrac{1}{2}(S-\Delta),\tfrac{1}{2}(S+\Delta))$, due to the fact that under BS transformation $p(S_r, \Delta_r|\tfrac{1}{2}(S-\Delta),\tfrac{1}{2}(S+\Delta))$ is proportional to the Kronecker delta $\delta_{S_r,S}$, simplifies to the following
\begin{align}
p^{S,\Delta}(\Delta_r) ={}&  \frac{1}{ 2^{S} (\tfrac{S-\Delta}{2})! (\tfrac{S+\Delta}{2})!}
\Bigg|\sum_{q=0}^{\tfrac{S-\Delta}{2}} \sum_{p=0}^{\tfrac{S+\Delta}{2}} \delta_{p+q,\tfrac{S-\delta}{2}}
\label{distribution}\\&
\binom{\tfrac{S-\Delta}{2}}{q} \binom{\tfrac{S+\Delta}{2}}{p}
(-1)^{p}
%\nonumber\\&
 \sqrt{(\tfrac{S-\Delta_r}{2})! (\tfrac{S+\Delta_r}{2})!}
\Bigg|^2.
\nonumber
\end{align}
The calculations that lead one to the formula closely resemble the ones presented in Appendix B, for a slightly more general process. 

\begin{figure}
  \begin{center}
    \raisebox{3cm}{(a)}
    \includegraphics[height=3.5cm]{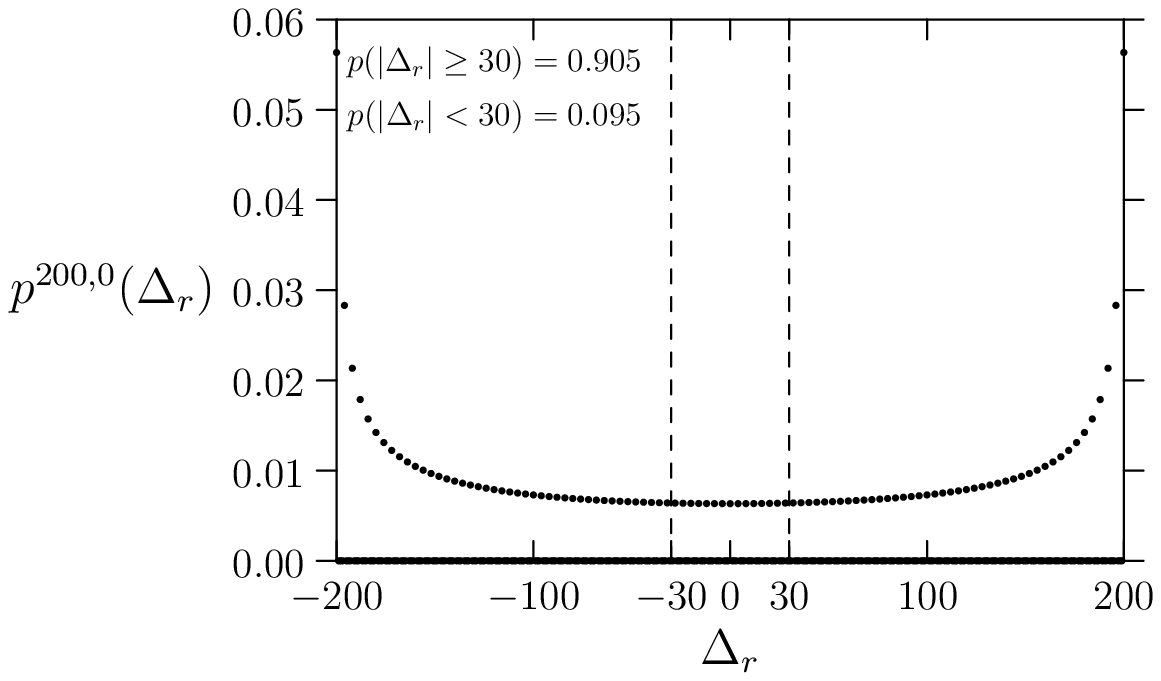}\\
    \raisebox{3cm}{(b)}
    \includegraphics[height=3.5cm]{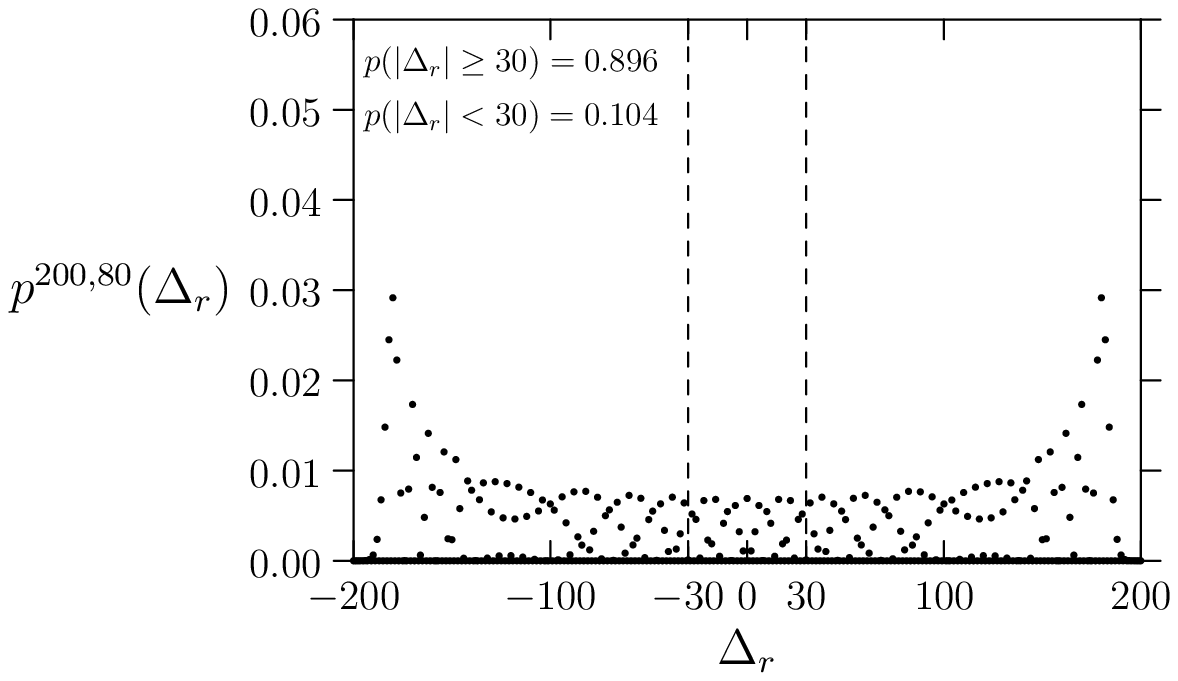}\\
    \raisebox{3cm}{(c)}
    \includegraphics[height=3.5cm]{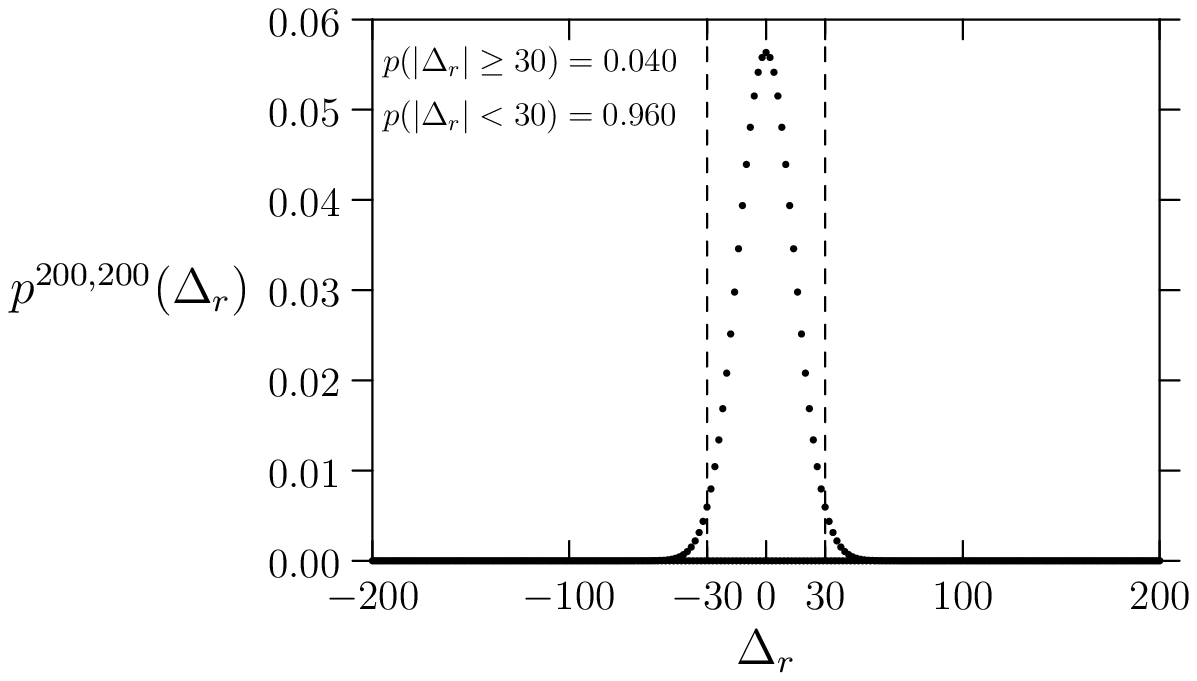}\\
  \end{center}
\caption{Distribution of the population difference $p^{S,\Delta}(\Delta_r)$ in a superposition Fock input state $\lvert\Psi_{in}\rangle =\lvert n,m\rangle_r$ conditioned on the measurement of $S=200$ photons and $\Delta = 0$ (a), $\Delta = 80$ (b), $\Delta = 200$ (c) at the PBS output. The vertical dashed lines show the threshold $\delta_{th}=30$. The probability that $|\Delta_r| \ge 30$ is given by $p(|\Delta_r| \ge 30)$.}
  \label{fig:single_mzi_s200}
\end{figure}

%We would like to emphasize that the particular shape of the superposition $\lvert\Psi_{in}\rangle$ set by the probability amplitudes $\xi_{n m}$ has no meaning for the filtering with the MDF, i.e. the MDF is state independent. These amplitudes give only the probability of a certain outcome from the filtering process, but do not influence $p^{S,\Delta}(\Delta_r)$.

The analysis of Eq.~(\ref{distribution}) shows that for a Fock state input with $|\Delta_r|\approx0$ one finds $|\Delta|\approx S$ with higher probability than $|\Delta|\approx 0$. {\it Vice versa}, when $|\Delta_r|\approx S$ the result $|\Delta|\approx 0$ is more likely than $|\Delta|\approx S$ \cite{campos}. Thus, the filter works probabilistically and for any outcome $S$ and $\Delta$ obtained all values of $\Delta_r$ are possible, but not equally probable. So we argue if $K$ and $L$ differ little, $\Delta \approx 0$, $|\Delta_r| \approx S$ is the most probable case, which means that a large initial population difference is anticipated. If $K$ and $L $ differ a lot, $\Delta \approx S$, we obtain that $|\Delta_r| \approx 0$ is favored and a small initial population difference has probably occurred. Figs.~\ref{fig:single_mzi_s200} depict the probability distribution $p^{S,\Delta}(\Delta_r)$ plotted for exemplary values of $S=200$, $\Delta=0$, $\Delta=80$ and $\Delta=200$.  The erratic shape of distributions in Figs.~\ref{fig:single_mzi_s200} reveals the interference between two non-zero Fock states entering a beamsplitter.

Imposing a filtering threshold in Eq.~(\ref{theoretical-preselection}) corresponds to fixing two independent threshold values. We choose a threshold value $\delta_{th}$ for which we check if $|\Delta_r| \ge \delta_{th}$. Next, since the process is probabilistic (is governed by the probability distribution $p^{S,\Delta}(\Delta_r)$), we fix the level of trust for it, i.e. the minimum probability, e.g. equal $90\%$, with which the condition $|\Delta_r| \ge \delta_{th}$ is fulfilled. The probability that the condition holds true is denoted by $p(|\Delta_r| \ge \delta_{th})$. It is evaluated by summing all probabilities $p^{S,\Delta}(\Delta_r)$ of these possibilities where $|\Delta_r| \ge \delta_{th}$, i.e. for $\Delta_r \in [ -S, -\delta_{th} ] \cup [ \delta_{th}, S ]$. Thus, if for a fixed value of $\delta_{th}$ $S$ increases, the probability $p(|\Delta_r| >= \delta_{th})$ increases as well. In Fig.~\ref{fig:single_mzi_s200} we set $\delta_{th}=30$. For $\Delta=200$, the probability of $|\Delta_r| \ge 30$ equals $p(|\Delta_r| \ge 30)=0.028 < 0.9$ and thus, this event is discarded. For $\Delta=0$, the probability is $p(|\Delta_r| \ge 30)=0.9$ and the event is accepted. 

%The above reasoning holds true also for inputs being Fock state superpositions. The necessary condition is to consider the value of $S_m$ and analyze the probability distribution $p^{S_m,\Delta_m}(\Delta_0)$ for each measurement outcome separately (the superposition components may involve different total photon numbers). In this case, the setup in Fig.~\ref{fig:MZI}a will filter out with high probability only those components of a superposition, whose occupation difference fulfill $|\Delta_0| \ge \delta_{th}$ and which have the total photon number equal to $S_m$. A detailed analysis of filtering of a Fock state superposition using the MDF is given in Appendix B.

In order to apply the MDF for the measurement induced operations, e.g. preparing the state for some further tasks, the whole setup must be like the one in Fig.~\ref{fig:MZI}b. A small portion of an incoming light is reflected (tapped) by a highly biased BS and examined by the scheme of Fig.~\ref{fig:MZI}a  located in a feed-forward loop. Since the reflected and transmitted beams are correlated, estimating the modulus of the population difference for the former gives an estimate for the latter. In this case, the MDF conditioned on the measurement outcome for the reflected beam, activates a shutter which passes or blocks the transmitted (almost unaffected by tapping) beam.  It is worth noting that the tapping relies on the fact that a polarization neutral BS splits the average intensities of both polarizations proportionally to its transmitivity $t$ and reflectivity $r\ll t$. This, in case of high photon numbers, means splitting with highest probability of photon numbers (of incoming two-mode Fock basis states) also in this proportion and that the initial ratio of occupations of the two polarization modes in a Fock component is preserved in the reflected and transmitted beams. 

We will illustrate the action of the tapping and the feed-forward loop from Fig.~\ref{fig:MZI}b using a Fock state $|\Psi_{in}\rangle = |n, m\rangle$ with an unknown initial population difference $\Delta_0=n-m$.  After the tapping BS, $v$ photons of $n$ are reflected from the first and $w$ photons of $m$ are reflected from the second input mode. The possible mode population differences equal $\Delta_r=v-w$ in the reflected beam and $\Delta_t = n - v - m + w$ in the transmitted beam, where $v \in [ 0, n ]$, $w \in [ 0, m ]$. The mode occupation difference registered at the detectors reads again $\Delta = L-K$. If the reflectivity of the tapping BS is $r=10\%$, the analysis of the probability distribution for the BS shows that for highly populated input $\Delta_r \simeq 0.1 \Delta_0$ and $\Delta_t \simeq 0.9 \Delta_0$. Now, the problem is reduced to the previously discussed: from the analysis below Eq.~(\ref{distribution}) we know, that if the measured in MDF $\Delta \simeq 0$, than entering MDF difference $\Delta_r$ and thus $\Delta_t$ are large; {\em vice versa}, if $\Delta$ is large, $\Delta_r \simeq 0$ and in consequence $\Delta_t \simeq 0$. In this setup, we directly set the threshold $\delta_{th}$ from Eq.~(\ref{theoretical-preselection}) for the transmitted beam, i.e. we require that $|\Delta_t| \ge \delta_{th}$, and the analysis of the reflected beam by MDF tells us the probability distribution of the population difference for the transmitted beam $p^{S,\Delta}(\Delta_t)$ and thus, the probability $p(|\Delta_t| \ge \delta_{th})$ with which this condition is fulfilled. Only if it is high enough, the MDF opens the shutter.

The above discussion applies also for Fock superposition states. See Appendix B for the complete calculus of the state evolution through the setup from Fig.~\ref{fig:MZI}b for an arbitrary superposition state and the derivation of the probability distribution of the population difference for the transmitted beam $p^{S,\Delta}(\Delta_t)$ (Eq.~\ref{Distribution}). 

Finally, we would like to mention that the assumption of the accurate measurement of $K$ and $L$ numbers is justified:  a setup involving losses after the tapping BS is equivalent to a setup with losses introduced in the reflected beam before the detectors. In the latter case, losses account for the imperfect detection. Thus, considering losses only in the transmitted part and perfect detection in the reflected part gives the full view. In experiments, a measurement accuracy of 150 photons, together with mean photon numbers per mode  $10^4$, would give a very good relative accuracy.

The discussion concerning weak invasibility of the MDF measurement on the beam leaving the shutter is moved to the Appendix C.

\section{Conclusions}

Thus, we have shown that the MDF is feasible and allows one to perform a threshold measurement while maintaining quantum superpositions. It works for any highly populated two-mode polarization states containing a single frequency and wavevector mode.   Realization of such a device is demanding, but the properties of the MDF are worth the effort.  The filter would be useful in the engineering of macroscopic quantum states of light. In the case of macro-qubits it circumvents the problem of inefficient detection, and  improves distinguishability. Thus, it makes them useful in quantum information and metrology protocols.

\vskip-3mm
\acknowledgments
\vskip-3mm

This work was supported by the EU FP7 Marie Curie Career Integration Grant No.\ 322150 "QCAT", MNiSW grant No.\ 2012/04/M/ST2/00789 and FNP Homing Plus project. MZ acknowledges EU Q-ESSENCE project, MNiSW (NCN) grant No.\ N202~208538. MC acknowledges EU FP7 BRISQ2 project (grant No.\ 308803). Calculations were performed at CI TASK in Gda\'nsk and Cyfronet in Krak\'ow.

\section*{Appendix A: Action of theoretical MDF on macro-qubits taking into account losses}

After filtering with the operator $\mathcal{P}_{\delta_{th}}$ the macro-qubits in Eq.~(\ref{macro-qubits}) take the form
\begin{eqnarray}
|\Phi\rangle &=& \sum_{\substack{i,j=0;\lvert 2i+1-2j \rvert\ge {\delta}_{th}}}^{\infty} \!\tilde{\gamma}_{ij}
\big|2i+1,2j\rangle,
\label{macro-qubits2}\\
|\Phi_{\perp}\rangle &=& \sum_{\substack{i,j=0;\lvert 2i+1-2j \rvert\ge {\delta}_{th}}}^{\infty} \!\tilde{\gamma}_{ij}
\big|2j,2i+1\rangle, \nonumber 
\end{eqnarray}
where the new probability amplitudes $\tilde{\gamma}_{ij}$ ensure the correct normalization. Next, the filtered macro-qubits are subjected to losses, modeled by a BS with the reflectivity $R$, which transforms them into mixed states 
\begin{align}
\rho_{\Phi}=& \sum_{\substack{i,j=0;\lvert 2i+1-2j \rvert\ge {\delta}_{th}}}^{\infty} \!\tilde{\gamma}_{ij}
\sum_{\substack{i',j'=0;\lvert 2i'+1-2j' \rvert\ge {\delta}_{th}}}^{\infty} \!\tilde{\gamma}_{i'j'}
\nonumber\\&\quad
\sum_{n=0}^{\min(2i+1,2i'+1)}\sum_{m=0}^{\min(2j,2j')} c_n^{(2i+1)} c_d^{(2j)} c_n^{(2i'+1)} c_d^{(2j')}
\nonumber\\&\quad
|2i+1-n, 2j-m\rangle\langle 2i'+1-n, 2j'-m|,
\\
\rho_{\Phi_{\perp}}=& \sum_{\substack{i,j=0;\lvert 2i+1-2j \rvert\ge {\delta}_{th}}}^{\infty} \!\tilde{\gamma}_{ij}
\sum_{\substack{i',j'=0;\lvert 2i'+1-2j' \rvert\ge {\delta}_{th}}}^{\infty} \!\tilde{\gamma}_{i'j'}
\nonumber\\&\quad
\sum_{n=0}^{\min(2i+1,2i'+1)}\sum_{m=0}^{\min(2j,2j')} c_n^{(2i+1)} c_d^{(2j)} c_n^{(2i'+1)} c_d^{(2j')}
\nonumber\\&\quad
|2j-m,2i+1-n\rangle\langle 2j'-m,2i'+1-n|,\nonumber
\end{align}
where $c_n^{(x)}=\sqrt{\binom{x}{n}\,R^n\,(1-R)^{x-n}}$ is the BS probability amplitude for the BS reflecting of $n$ from $x$ photons.

The photon number distribution for these states is
\begin{align}
p_{\Phi}(k,l) &= \mathrm{Tr}\{\rho_{\Phi} |k,l\rangle\langle k,l| \},\\
p_{\Phi_{\perp}}(k,l) &= \mathrm{Tr}\{\rho_{\Phi_{\perp}} |k,l\rangle\langle k,l| \}, \nonumber
\end{align}

\begin{align}
p_{\Phi}(k,l) &= \sum_{\substack{i,j=0;\lvert 2i+1-2j \rvert\ge {\delta}_{th}}}^{\infty} \tilde{\gamma}_{ij}^2
\left(c_{2i+1-k}^{(2i+1)}\right)^2 \left(c_{2j-l}^{(2j)}\right)^2 \nonumber\\
&\quad \Theta(2i+1-k) \Theta(2j-l), \label{eq:phnum} \\
p_{\Phi_{\perp}}(k,l) &= \sum_{\substack{i,j=0;\lvert 2i+1-2j \rvert\ge {\delta}_{th}}}^{\infty} \tilde{\gamma}_{ij}^2
\left(c_{2i+1-l}^{(2i+1)}\right)^2 \left(c_{2j-k}^{(2j)}\right)^2 \nonumber\\
&\quad \Theta(2i+1-l) \Theta(2j-k), \nonumber
\end{align}
where $\Theta(x)=1 (0)$ for $x \ge 0$ ($x<0$).

\section*{Appendix B: MDF measurement and the state evolution in tapping and feed-forward loop}

\begin{figure}
\begin{center}
  \includegraphics[width=7cm]{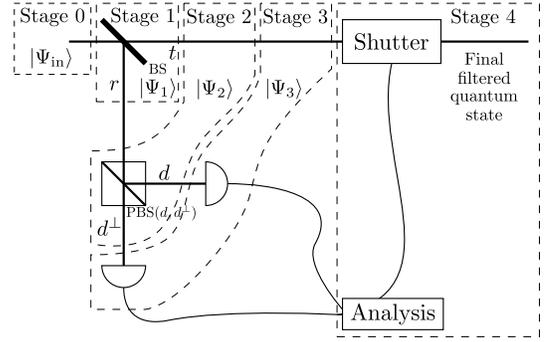}
\end{center}
\caption{Physical implementation of the MDF with the notation indicating the state evolution in different parts of the setup.}
\label{schema_D}
\end{figure}

In this appendix we will present the evolution of an input state $\lvert\Psi_{in}\rangle =\sum_{n,m}\xi_{n m}\lvert n,m\rangle$ entering the setup depicted in Fig.~\ref{fig:MZI}b. 

Fig.~\ref{schema_D} illustrates each stage of the experiment performed by this setup. At stage 1 this state impinges on a tapping BS, with the reflectivity coefficient $r$, which acts independently on both polarization modes. This results in transformation $\mathcal{U}_{\text{BS}}\lvert n,m\rangle$. Its action on a single polarization Fock state reads
\begin{align}
\mathcal{U}_{\text{BS}}\lvert0,n\rangle &=\sum_{v=0}^n c_v^{(n)}\lvert v\rangle_r \lvert n-v\rangle_t,
\\
c_v^{(n)} &=\sqrt{\binom{n}{v}\,r^v\,(1-r)^{n-v}}.
\nonumber
\end{align}
The index $r$ ($t$) corresponds to the reflected (transmitted) part. The input state is transformed to $\lvert\Psi_{1}\rangle = \mathcal{U}_{\text{BS}}\lvert\Psi_{in}\rangle$ where 
\begin{align}
& \lvert\Psi_{1}\rangle
= \sum_{n,m}\xi_{nm} \sum_{v=0}^n \sum_{w=0}^m c_v^{(n)}\, c_w^{(m)}
\, \lvert v,w\rangle_r \lvert n-v,m-w\rangle_t.
\end{align}
Next, in stage 2, the reflected beam impinges on the PBS. It transforms the operators $a_r$ and $a_{r^{\perp}}$ according to the transformation
$a_d=1/\sqrt{2}(a_r+a_{r^{\perp}})$, $a_{d^{\perp}}=1/\sqrt{2}(a_{r^{\perp}}-a_r)$. The reflected part $\lvert v, w\rangle_{r}=\tfrac{1}{\sqrt{v!\,w!}}\, (a_r^{\dagger})^v\, (a_{r^{\perp}}^{\dagger})^w$ looks as follows
\begin{align}
& \mathcal{U}_{\text{PBS}} \lvert v, w\rangle_{r}
= \tfrac{1}{\sqrt{v!\, w!}} \tfrac{1}{\sqrt{2^{v+w}}}
\,(a_d^{\dagger}-a_{d^{\perp}}^{\dagger})^v\, (a_d^{\dagger}+a_{d^{\perp}}^{\dagger})^w\, \lvert0\rangle \nonumber
\\
&= \tfrac{1}{\sqrt{v!\, w!}} \tfrac{1}{\sqrt{2^{v+w}}}
\sum_{p=0}^v \sum_{q=0}^w \binom{v}{p} \binom{w}{q}
(a_d^{\dagger})^p\,(-a_{d^{\perp}}^{\dagger})^{v-p}\nonumber\\&\quad\quad\quad (a_d^{\dagger})^q\,(a_{d^{\perp}}^{\dagger})^{w-q}
\,\lvert0\rangle \nonumber
\\
&= \tfrac{1}{\sqrt{v!\, w!}} \tfrac{1}{\sqrt{2^{v+w}}}
\sum_{p=0}^v \sum_{q=0}^w \binom{v}{p} \binom{w}{q} (-1)^{v-p}\nonumber\\&\quad\quad\quad
(a_d^{\dagger})^{p+q}\,(a_{d^{\perp}}^{\dagger})^{v+w-p-q}\, \lvert0\rangle.
\end{align}

After the PBS the state equals $\lvert\Psi_{2} \rangle= \mathcal{U}_{\text{PBS}} \mathcal{U}_{\text{BS}}\lvert\Psi_{in}\rangle$
\begin{align}
& \lvert\Psi_{2}\rangle
= \sum_{n,m}\xi_{nm} \sum_{v=0}^n \sum_{w=0}^m
\dfrac{c_v^{(n)}\, c_w^{(m)}}{\sqrt{v!\, w!}}
\tfrac{1}{\sqrt{2^{v+w}}}
\nonumber\\&\quad
\sum_{p=0}^v \sum_{q=0}^w \binom{v}{p} \binom{w}{q} (-1)^{v-p}
\sqrt{(p+q)!\,(v+w-p-q)!}
\nonumber\\&\quad \lvert p+q,v+w-p-q\rangle_d
 \lvert n-v,m-w\rangle_t.
\end{align}

In stage 3 the detectors detect two Fock states $|K,L\rangle_d$ and project the state $\lvert\Psi_{2} \rangle$ to $\lvert\Psi_{3} \rangle = {}_d\langle K,L\rvert \mathcal{U}_{\text{PBS}} \mathcal{U}_{\text{BS}}\lvert\Psi_{\text{in}}\rangle$

\begin{align}
& \lvert\Psi_{\text{3}}\rangle
= \sum_{n,m}\tilde{\xi}_{nm} \sum_{v=0}^n \sum_{w=0}^m 
\dfrac{c_v^{(n)}\, c_w^{(m)}}{\sqrt{v!\,w!}}
\tfrac{1}{\sqrt{2^{v+w}}}
\nonumber\\&\quad
\sum_{p=0}^v \sum_{q=0}^w \binom{v}{p} \binom{w}{q} (-1)^{v-p}
\sqrt{(p+q)!\,(v+w-p-q)!}
\nonumber\\&\quad 
\delta_{K,p+q}\,\delta_{L,v+w-p-q}\, \lvert n-v,m-w\rangle_t
\nonumber
\\
&= \sum_{n,m}\tilde{\xi}_{nm} \sum_{v=0}^n \sum_{w=0}^m 
\dfrac{c_v^{(n)}\, c_w^{(m)}}{\sqrt{v!\,w!}}
\tfrac{1}{\sqrt{2^{v+w}}}
\sum_{p=0}^v \sum_{q=0}^w \binom{v}{p} \binom{w}{q} 
\nonumber\\&\quad
(-1)^{v-p} \sqrt{K!\,L!}
\,\delta_{K,p+q}\,\delta_{L,v+w-K}\, \lvert n-v,m-w\rangle_t.
\end{align}
The coefficients $\tilde{\xi}_{nm}$ are renormalized to ensure normalization of $\lvert\Psi_{3} \rangle$.

For the further discussion of the filtering process it is useful to compute the conditional photon number distribution for the transmitted beam
$p^{K,L}(k,l) = |\langle k,l\lvert\Psi_{3} \rangle |^2$
\begin{align}
&p^{K,L}(k,l)= K!\,L! \Big(\sum_{n,m} \tilde{\xi}_{nm} \sum_{v=0}^n \sum_{w=0}^m 
\dfrac{c_v^{(n)}\, c_w^{(m)}}{\sqrt{v!\,w!}}
\tfrac{1}{\sqrt{2^{v+w}}} 
\nonumber\\& \quad
\delta_{L,v+w-K} \delta_{k,n-v}\,\delta_{l,m-w} 
\nonumber\\& \quad
\sum_{p=0}^v \sum_{q=0}^w \binom{v}{p} \binom{w}{q} (-1)^{v-p}
\,\delta_{K,p+q}\Big)^2.
\label{pkl}
\end{align}
We change the variables $L$ and $K$ so that they were corresponding to the quantities useful for the filtering: the total sum of the registered photons $S = L+K$ and the difference in the occupation of the polarization modes $\Delta = L-K$. We obtain $p^{S,\Delta}(S_t,\Delta_t)$ with $S_t = k+l$, $\Delta_t = k-l$
\begin{align}
&p^{S,\Delta}(S_t,\Delta_t) =
\left(\tfrac{S+\Delta}{2}\right)!\,\left(\tfrac{S-\Delta}{2}\right)!
\nonumber\\& \quad
\Big(\sum_{n,m} \tilde{\xi}_{nm} \sum_{v=0}^n \sum_{w=0}^m 
\dfrac{c_v^{(n)}\, c_w^{(m)}}{\sqrt{v!\,w!}}
\tfrac{1}{\sqrt{2^{v+w}}} 
\nonumber\\& \quad
\delta_{\tfrac{S+\Delta}{2},v+w-\tfrac{S-\Delta}{2}}\, \delta_{\tfrac{S_t+\Delta_t}{2},n-v}\,\delta_{\tfrac{S_t-\Delta_t}{2},m-w} 
\nonumber\\& \quad
\sum_{p=0}^v \sum_{q=0}^w \binom{v}{p} \binom{w}{q} (-1)^{v-p}
\,\delta_{\tfrac{S-\Delta}{2},p+q}\Big)^2.
\end{align}
The probability distribution for the occupation difference in the transmitted beam $\Delta_t$ is given by
\begin{align}
p^{S,\Delta}(\Delta_t) = \sum_{S_t=0}^{\infty} p^{S,\Delta}(S_t,\Delta_t).
\label{Distribution}
\end{align}

The filtering is performed in stage 4 of the experiment. Here, the detectors' readings are analyzed and only those events and realizations of $\lvert\Psi_{\text{3}}\rangle$ are accepted where $\Delta \simeq 0$. Depending on the result of measurement of $\Delta$, the shutter is opened or remains closed and the state is rejected.

\begin{figure}[ht]
\vskip-1mm
\begin{center}
\raisebox{3.5cm}{(a)}
\includegraphics[width=7cm]{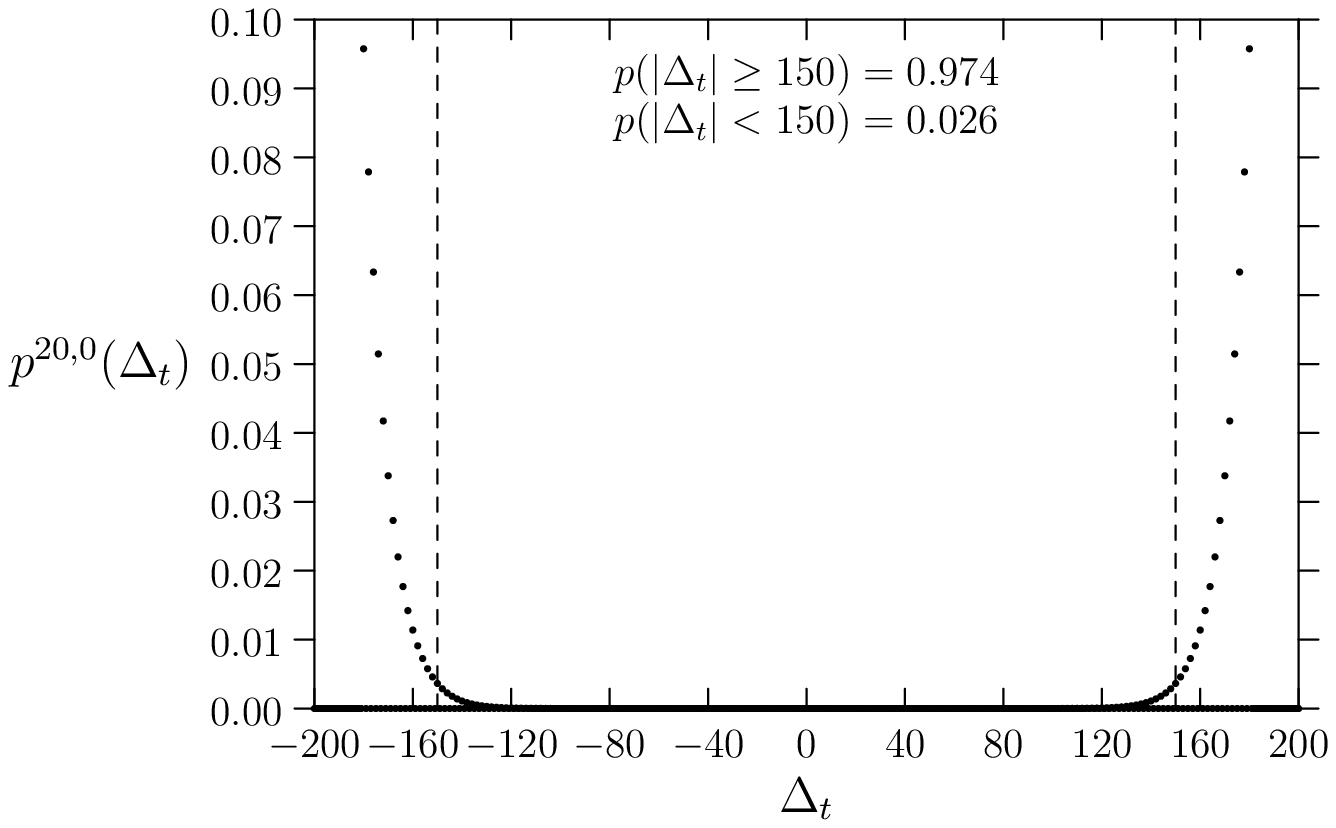}\\
\raisebox{3.5cm}{(b)}
\includegraphics[width=7cm]{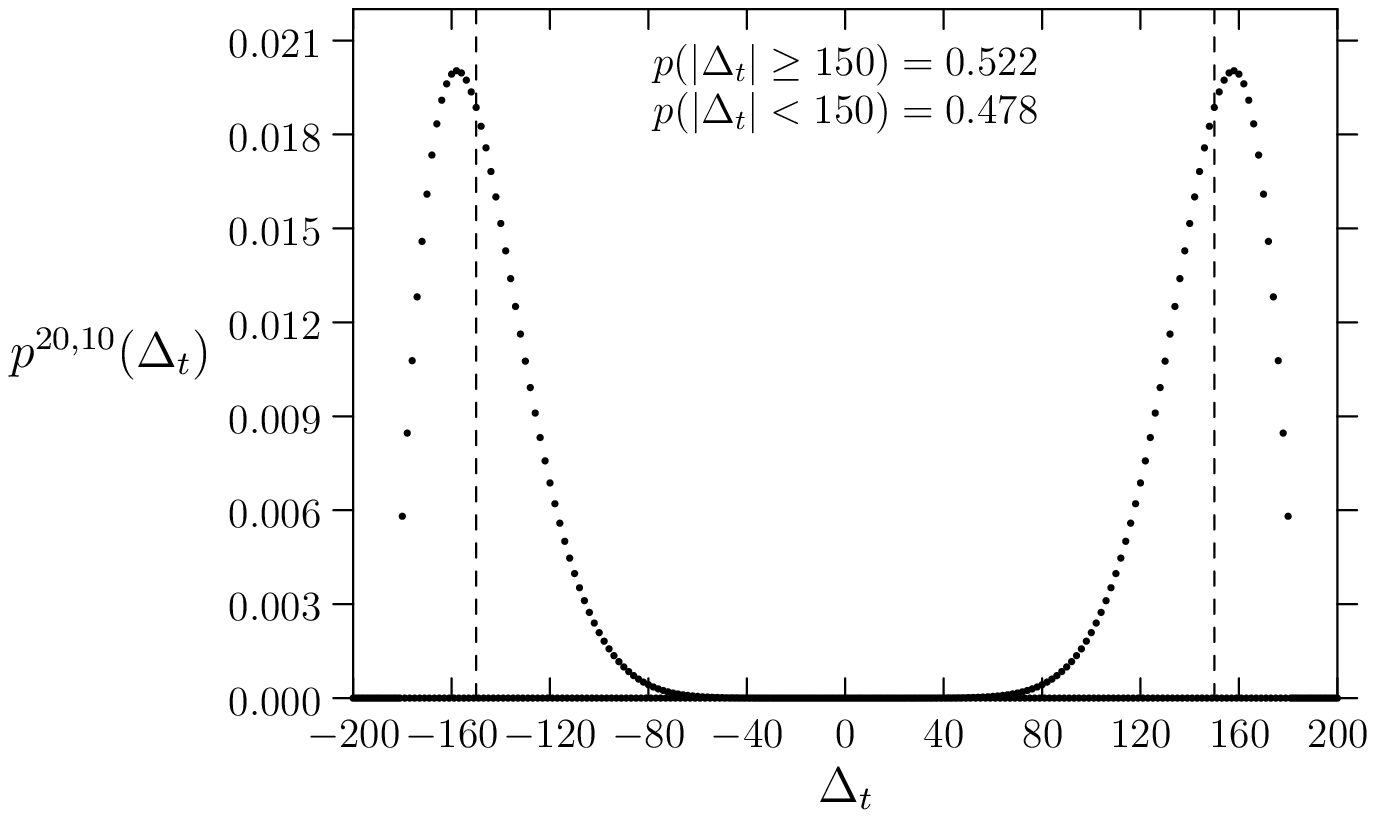}\\
\raisebox{3.5cm}{(c)}
\includegraphics[width=7cm]{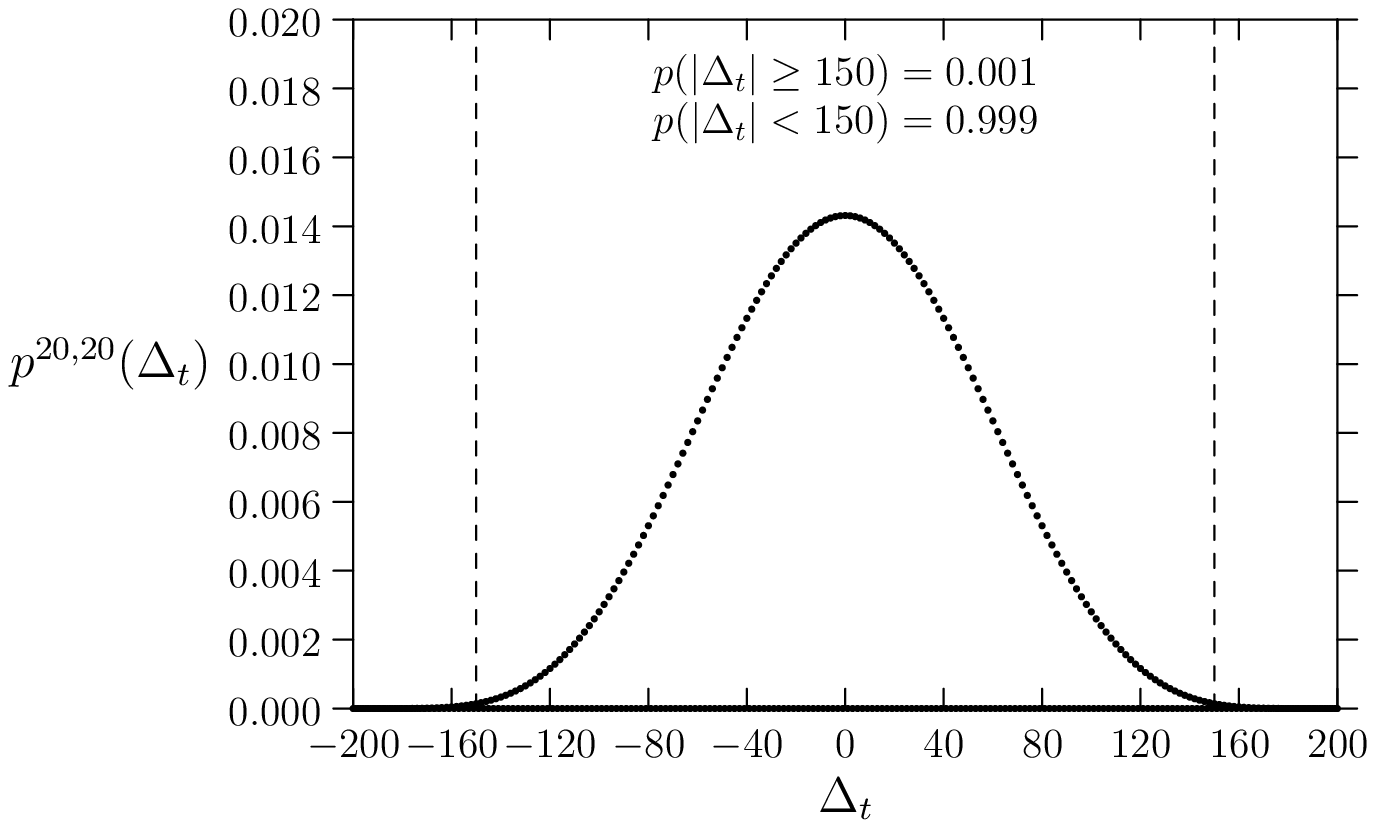}
\end{center}
\vskip-5mm
\caption{Distribution of the population difference $p^{S,\Delta}(\Delta_t)$ in the transmitted beam $t$ after the shutter for the state in Eq.~(\ref{example2}) with $S_0=200$ assuming that $S=20$ photons were registered in the reflected beam and the difference measured by detectors was $\Delta=0$ (a), $\Delta=10$ (b), $\Delta=20$ (c). The vertical dashed lines show the threshold $\delta_{th}=150$. The probability that $|\Delta_t| \ge 150$ is given by $p(|\Delta_t| \ge 150)$.}
\label{ProbabilityDistributions}
\vskip-5mm
\end{figure}

\subsubsection*{Example}
\vskip-1ex

We consider a simple superposition of Fock states with fixed total photon number $S_0$ (it allows avoiding the summation over $S_t$ in Eq.~(\ref{Distribution})) and with a uniform distribution of the occupation difference $\Delta_0$
\begin{equation}
\lvert\Psi_{in}\rangle = 1/\sqrt{S_0+1}\sum_{n=0}^{S_0} \lvert n,S_0-n \rangle.
\label{example2}
\end{equation}
In Fig.~\ref{ProbabilityDistributions} we have depicted the probability distributions $p^{S,\Delta}(\Delta_t)$ for this state with $S_0=200$ for three cases: $\Delta=0$, $\Delta=10$ and $\Delta=20$ for $S=20$. These plots reveal that for small $\Delta \approx 0 $ the most probable values of $\Delta_t$ in the transmitted beam are large. The higher $\Delta$ is, the more probable are the superposition components with $\Delta_t=0$ to be present in the output beam. We took $\delta_{th} =150$ and the probabilities that $|\Delta_t| \ge 150$ equal: $0.974$, $0.522$, $0.001$ for $\Delta=0$, $\Delta=10$, $\Delta=20$, respectively.

\section*{Appendix C: Small disturbance by MDF measurement of "macroscopic" qubits} \label{sec5}

In reality, one would aim at applying the MDF to more complex quantum states, the superpositions like the one given in Eq.~(\ref{macro-qubits}), which constitute a ``macroscopic'' qubit. The goal of the MDF apart form filtering of those states and increasing their distinguishability in classical detection, is to avoid discriminating between them. Moreover, usually the experimental conditions are not perfect and in the analysis of the action of the filter one has to take into account the multi-mode character of the input state and the losses. We will discuss these issues in this section.

\begin{figure}[b]
\begin{center}
\raisebox{3.5cm}{(a)}
\includegraphics[width=7cm]{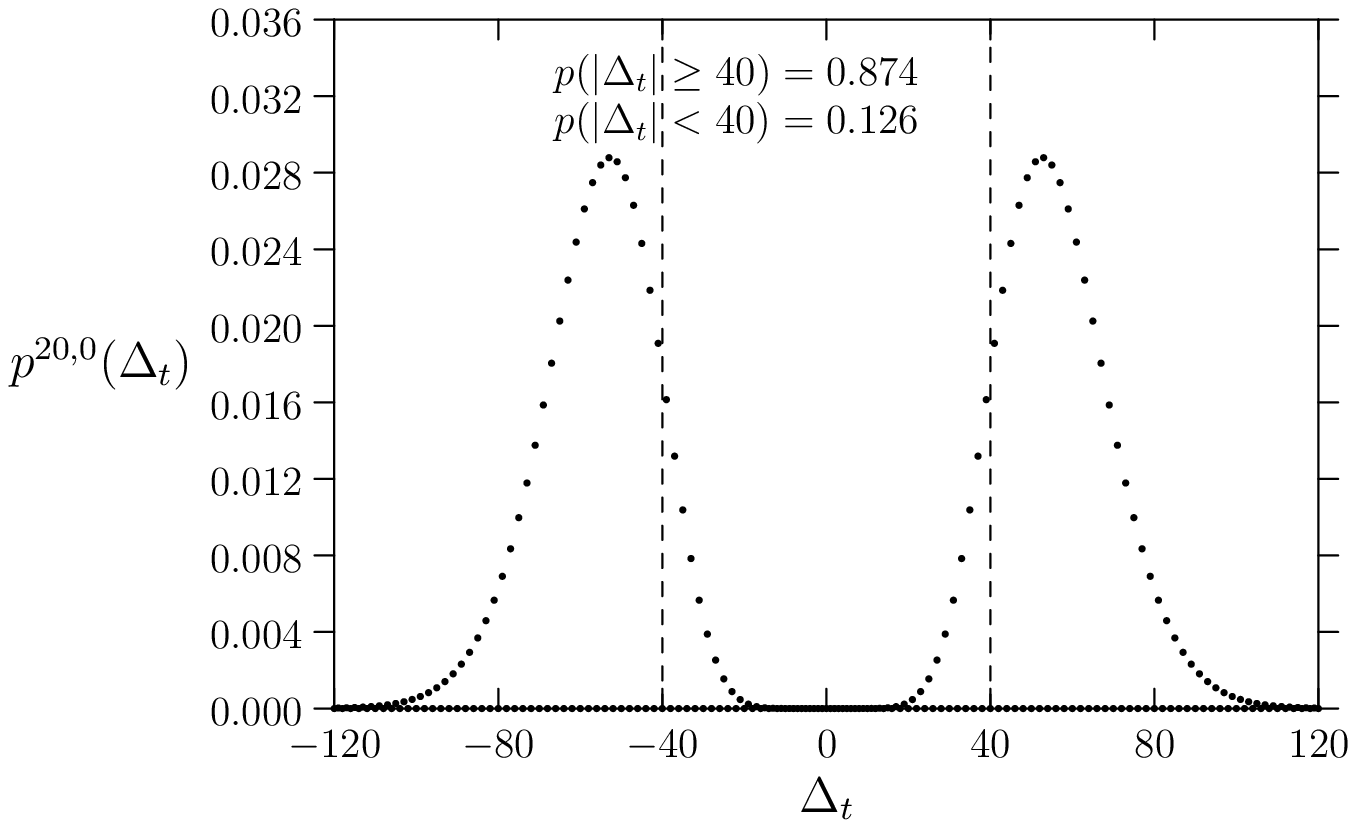}\\
\raisebox{3.5cm}{(b)}
\includegraphics[width=7cm]{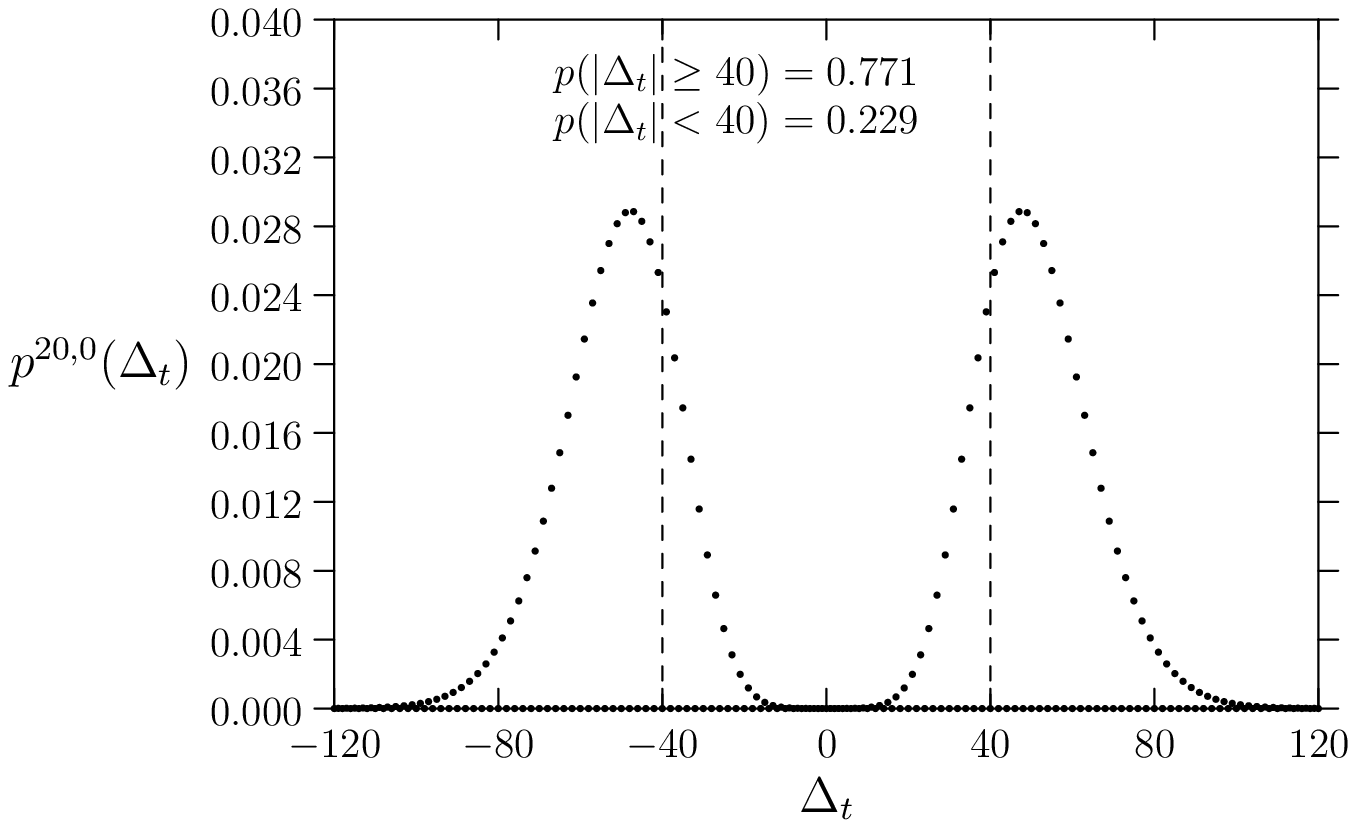}\\
\raisebox{3.5cm}{(c)}
\includegraphics[width=7cm]{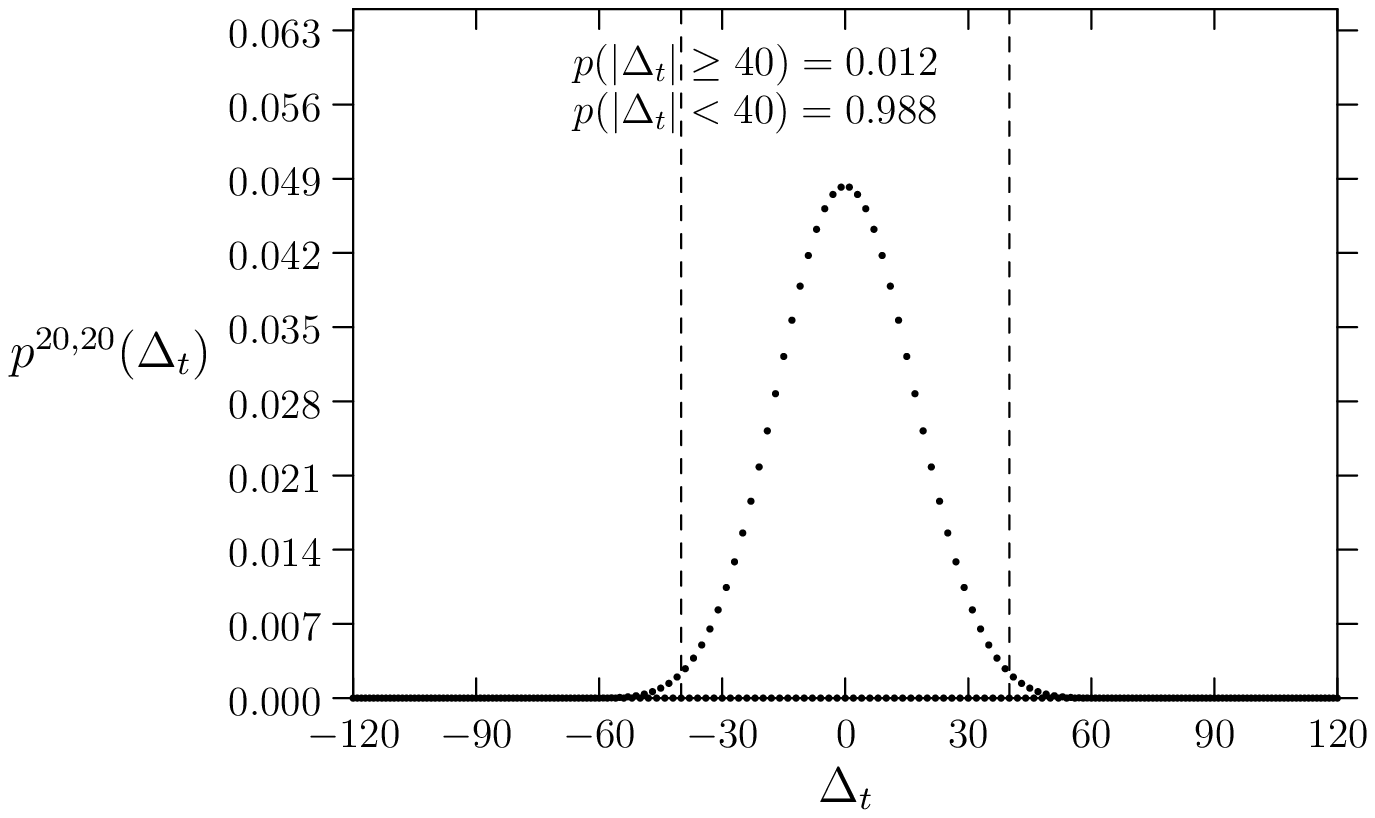}
\end{center}
\vskip-5mm
\caption{Distribution of the population difference $p^{S,\Delta}(\Delta_t)$ (Eq.~\ref{Distribution}) in the transmitted beam $t$ after the shutter for $\rho_{in} = 1/2(|\Phi\rangle \langle\Phi| + |\Phi_{\perp}\rangle \langle \Phi_{\perp}|)$ for $g=1.87$ assuming that $S=20$ photons were registered in the reflected beam and the difference measured by detectors was $\Delta=0$ (a), $\Delta=10$ (b), $\Delta=20$ (c). The vertical dashed lines show the threshold $\delta_{th}=40$. The probability that $|\Delta_t| \ge 40$ is given by $p(|\Delta_t| \ge 40)$.}
\label{ProbabilityDistributionsDeM}
\vskip-5mm
\end{figure}

\begin{figure}[b]
\begin{center}
\begin{tabular}{c}
$v=0.72$\\[-0.1cm]
\raisebox{3cm}{a)}\includegraphics[height=4cm]{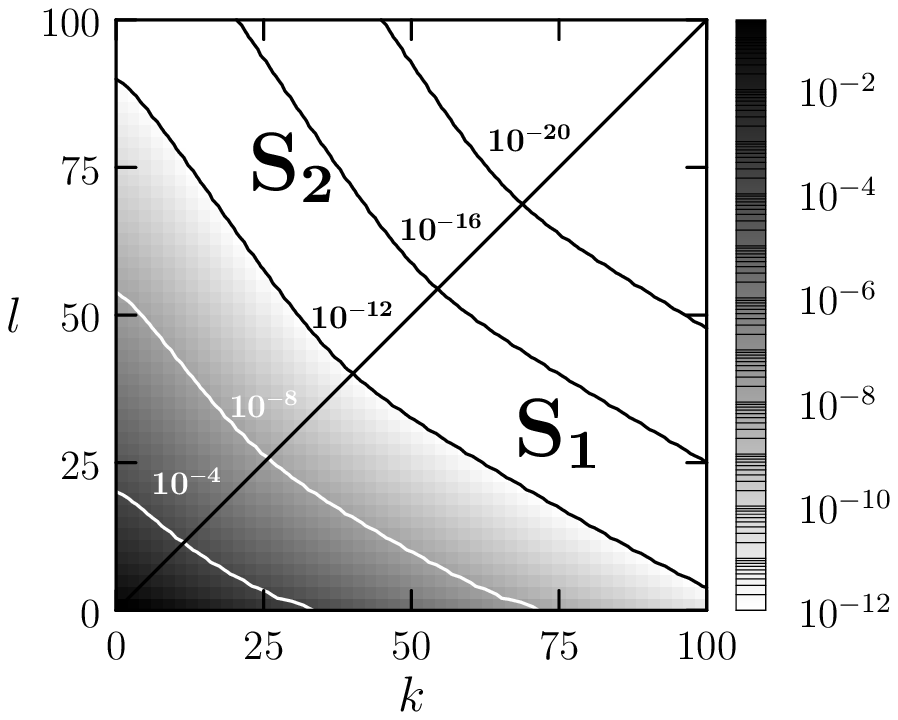}\\
$v=0.93$\\[-0.1cm]
\raisebox{3cm}{b)}\includegraphics[height=4cm]{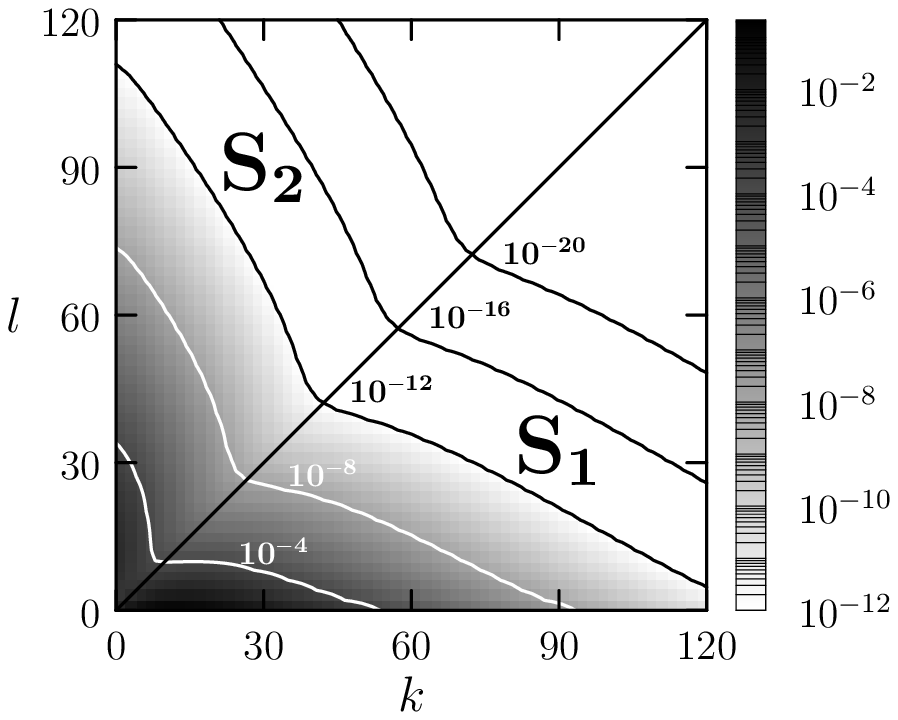}\\
$v=0.96$\\[-0.1cm]
\raisebox{3cm}{c)}\includegraphics[height=4cm]{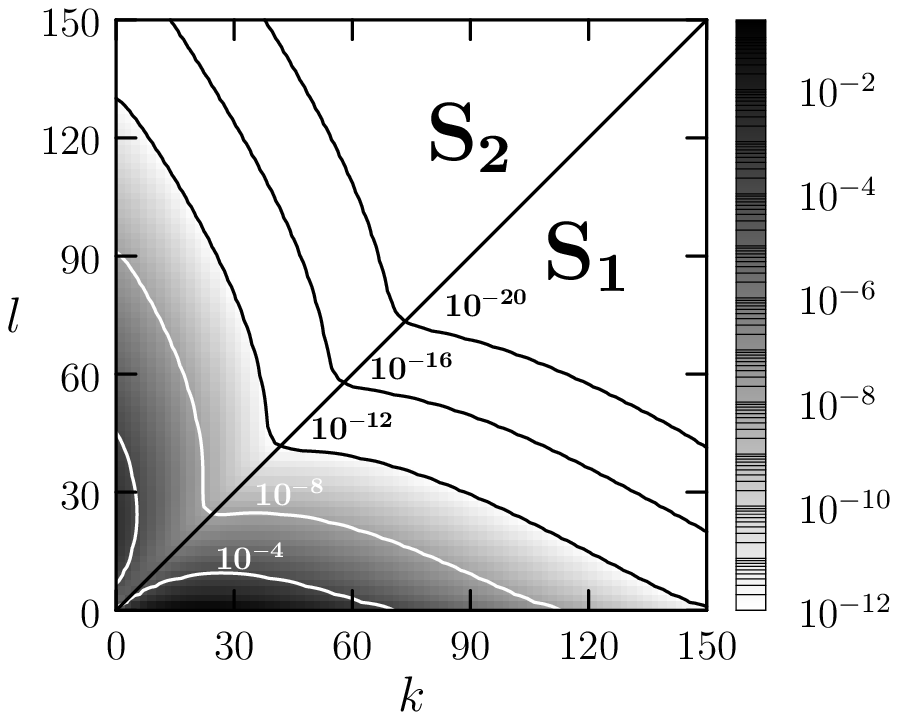}\\
$v=0.97$\\[-0.1cm]
\raisebox{3cm}{d)}\includegraphics[height=4cm]{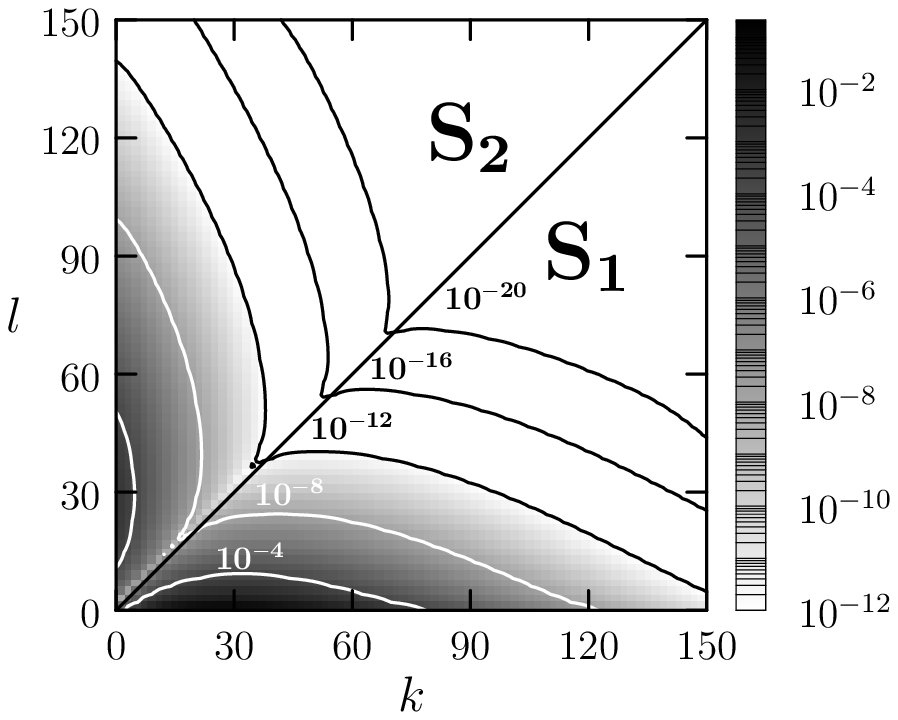}
\end{tabular}
\end{center}
\vskip-7mm
\caption{Photon number distribution $p_{\Phi}$ (Eq.~(\ref{eq:photon_number_distr_Phi})) and distinguishability $v$ (Eq.~(4)) of the macroscopic state $\lvert\Phi\rangle$ processed by the setup from Fig.~\ref{fig:MZI}b, computed for $g=1.87$, the level of trust $90\%$ and $\delta_{th}=0$ (a), $\delta_{th}=5$ (b), $\delta_{th}=10$ (c), $\delta_{th}=15$ (d). $k$ and $l$ denote numbers of photons in two orthogonal polarization modes.}
  \label{MacroPhotonNumber}
\vskip-5mm
\end{figure}

Imagine a source producing a micro-macro polarization singlet state of the form $|\Psi^-\rangle = (|1\rangle_A |\Phi_{\perp}\rangle_B - |1_{\perp}\rangle_A |\Phi\rangle_B)/\sqrt{2}$. The macroscopic part $B $ of the singlet is  fed to the setup in Fig.~\ref{fig:MZI}b. The initial state  reads 
\begin{equation}
\rho_{in} = 1/2(|\Phi\rangle \langle\Phi| + |\Phi_{\perp}\rangle \langle \Phi_{\perp}|).
\label{example3}
\end{equation}
The state passes through the whole setup in Fig.~\ref{fig:MZI}b. In Fig.~\ref{ProbabilityDistributionsDeM} we depicted the probability distributions $p^{S,\Delta}(\Delta_t)$ (Eq.~(\ref{Distribution}) with $\tilde{\xi}_{nm}=\tilde{\gamma}_{nm}$) for this state as a function of the population difference $\Delta_t$ in the transmitted beam $t$ after the shutter. In our computation we assumed the gain $g=1.87$, $S=20$ photons registered in the reflected beam and chose $\delta_{th} =40$. The probabilities $p(|\Delta_t| \ge 40)$ that $|\Delta_t| \ge 40$ are: $0.87$, $0.77$, $0.01$ for $\Delta=0$, $\Delta=10$, $\Delta=20$, respectively.

We also computed the photon number distributions (useful for the distinguishability estimation) for $\rho_{in}$ processed by the setup in Fig.~\ref{fig:MZI}b and compared them with the distributions obtained in theoretical filtering performed by $\mathcal{P}_{\delta_{th}}$ which are displayed in Fig.~\ref{fig:qfuncIdeal}. The photon number distribution for $\rho_{in}$ reads
\begin{align}
  p_{\Phi}(k,l)=\sum_{S\in\mathbf{S}}\,p^{S,\Delta=0}(k,l),
  \label{eq:photon_number_distr_Phi}
\end{align}

\begin{figure}[b]
\begin{center}
\raisebox{3.5cm}{(a)}
\includegraphics[width=7cm]{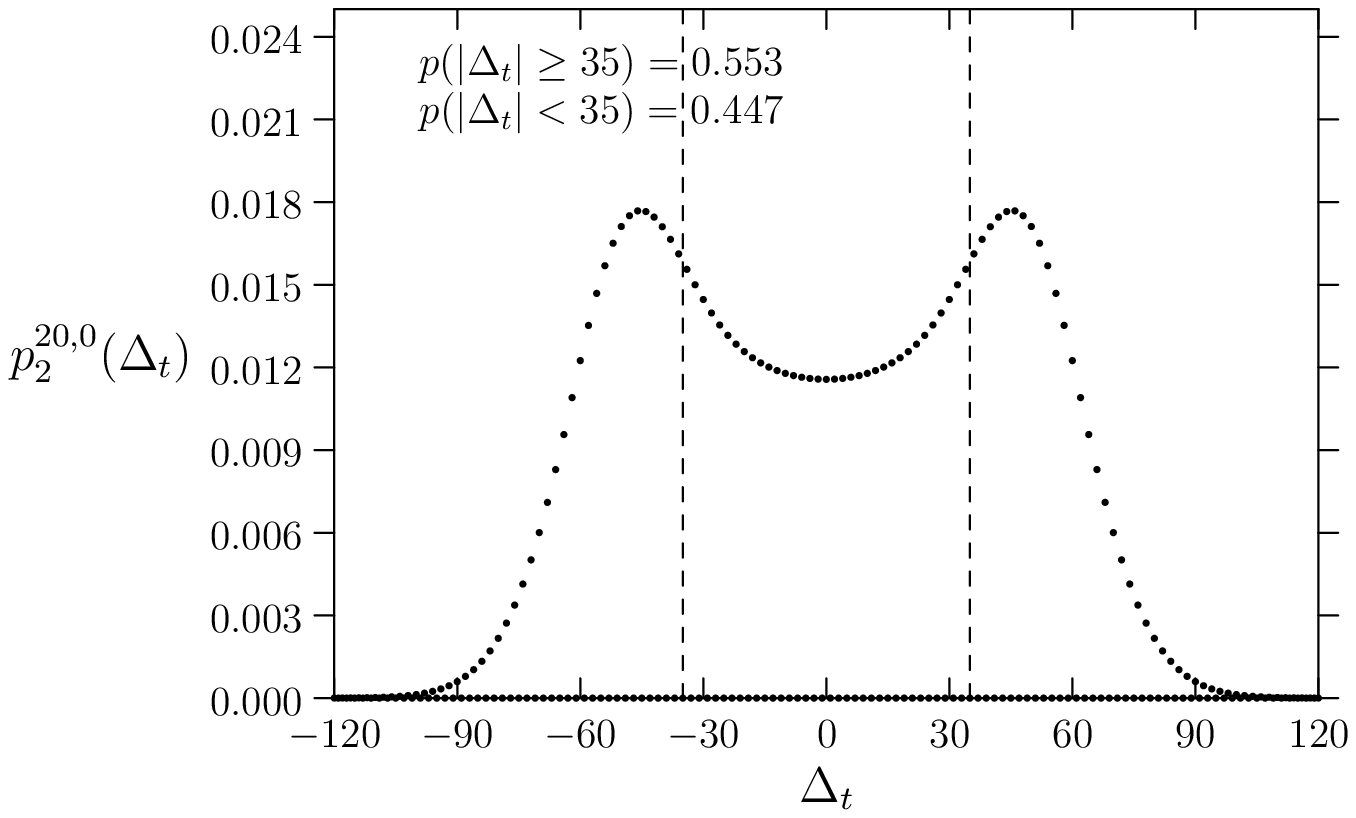}\\
\raisebox{3.5cm}{(b)}
\includegraphics[width=7cm]{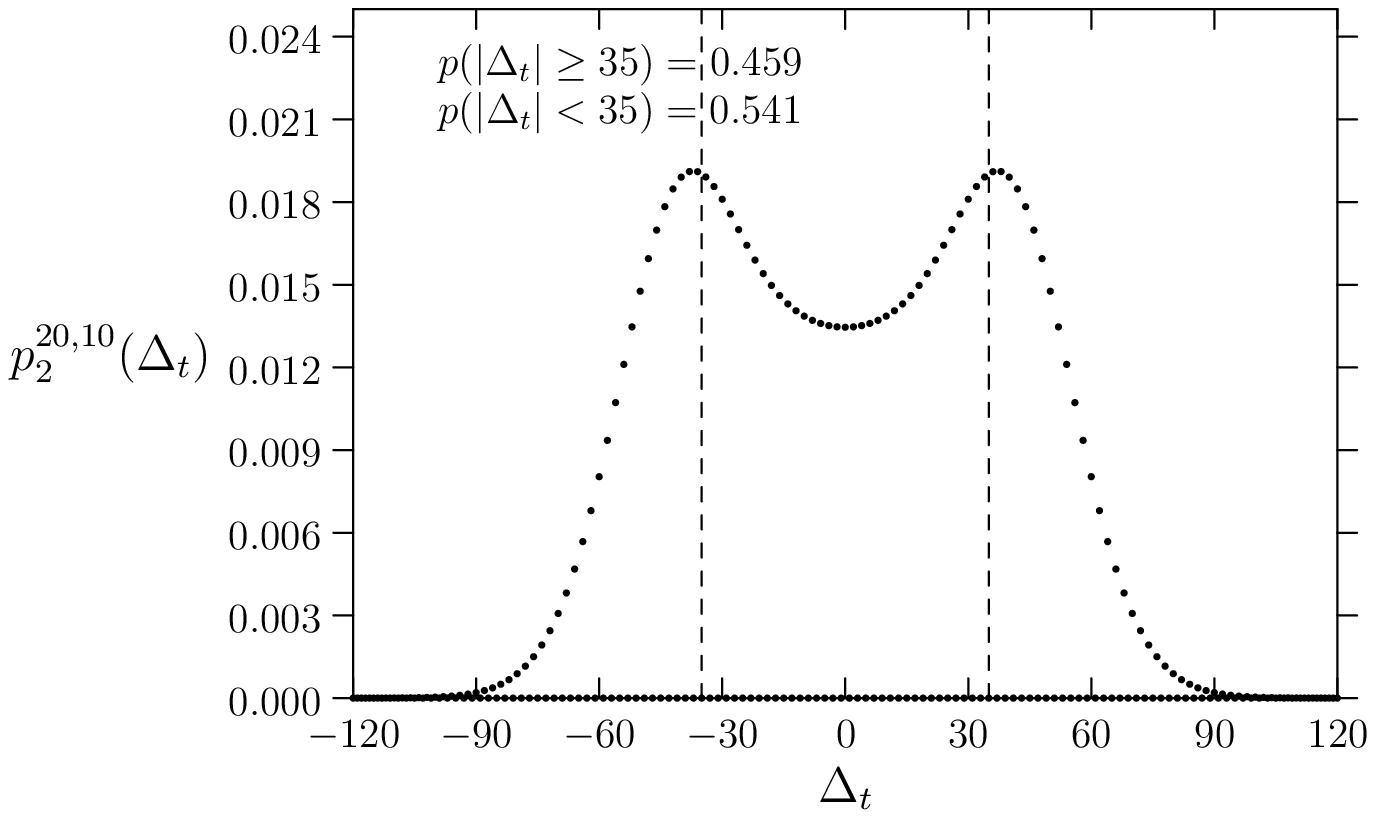}\\
\raisebox{3.5cm}{(c)}
\includegraphics[width=7cm]{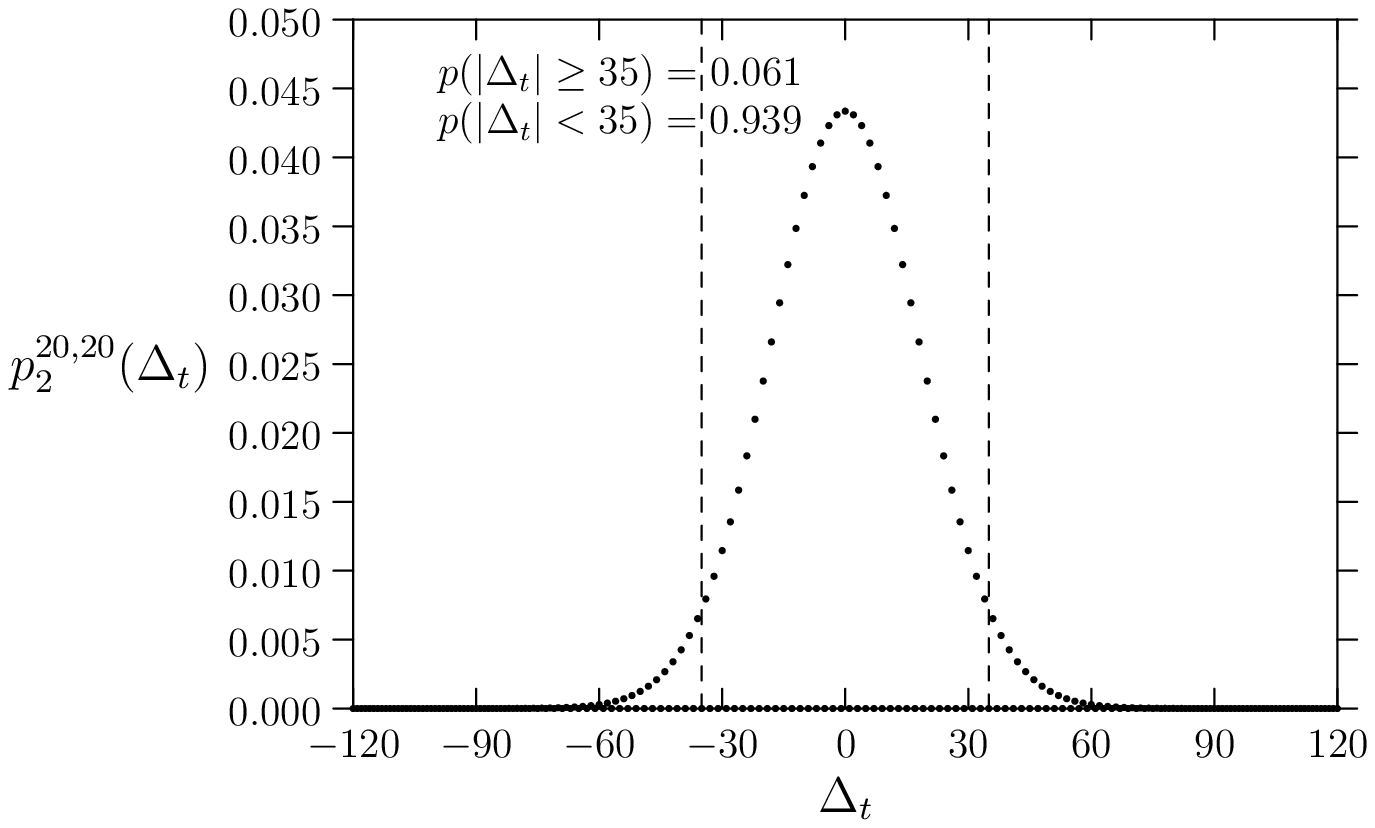}
\end{center}
\vskip-5mm
\caption{Distribution of the population difference $p^{S,\Delta}_2(\Delta_t)$ (Eq.~(\ref{DistributionMode})) in the transmitted beam $t$ after the shutter for the two-mode state in Eq.~(\ref{example3}) for $g=1.87$ assuming that $S=20$ photons were registered in the reflected beam and the difference measured by detectors was $\Delta=0$ (a), $\Delta=10$ (b), $\Delta=20$ (c).}
\label{ProbabilityDistributionsMultiMode}
\vskip-5mm
\end{figure}

\begin{figure}[b]
\begin{center}
\vskip-1mm
\raisebox{3.5cm}{(a)}
\includegraphics[width=7cm]{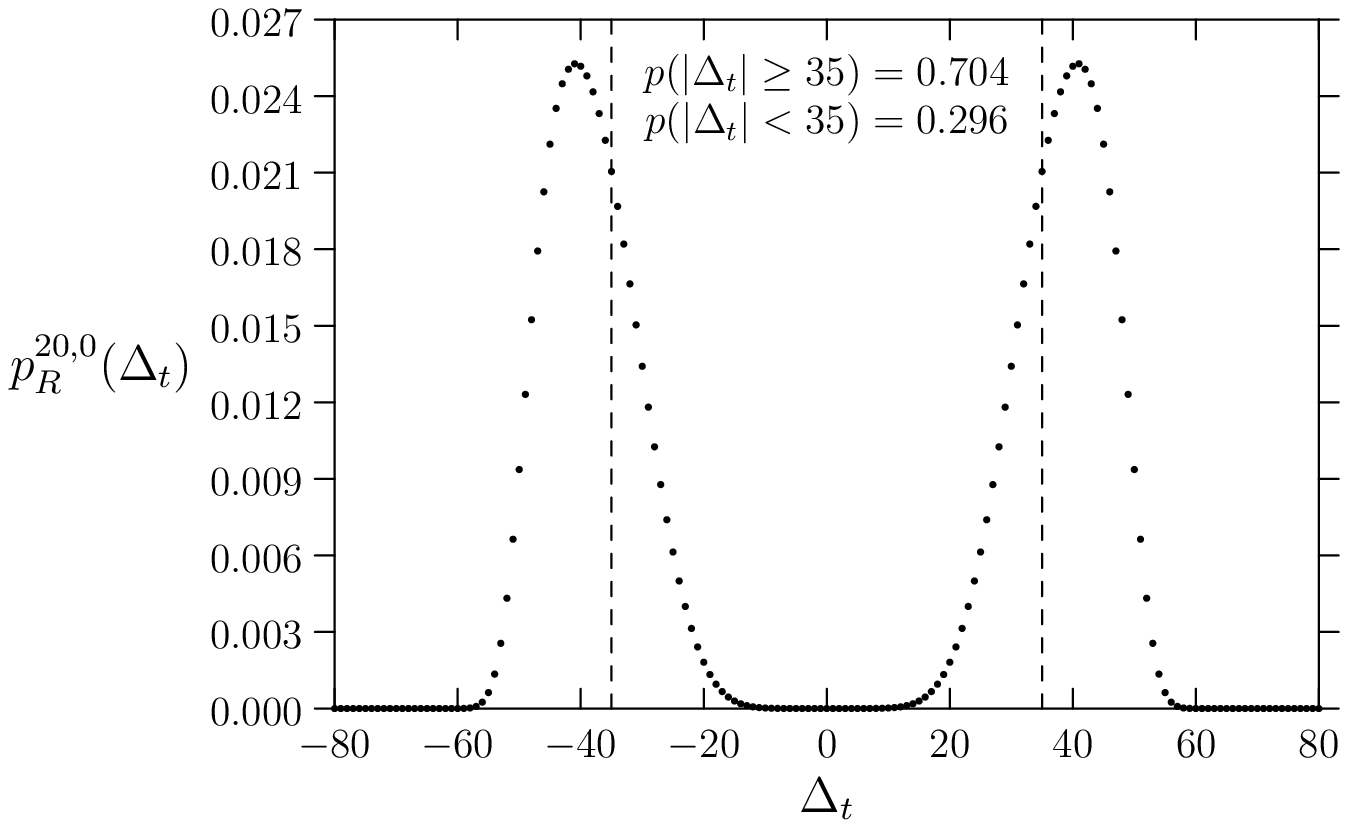}\\
\raisebox{3.5cm}{(b)}
\includegraphics[width=7cm]{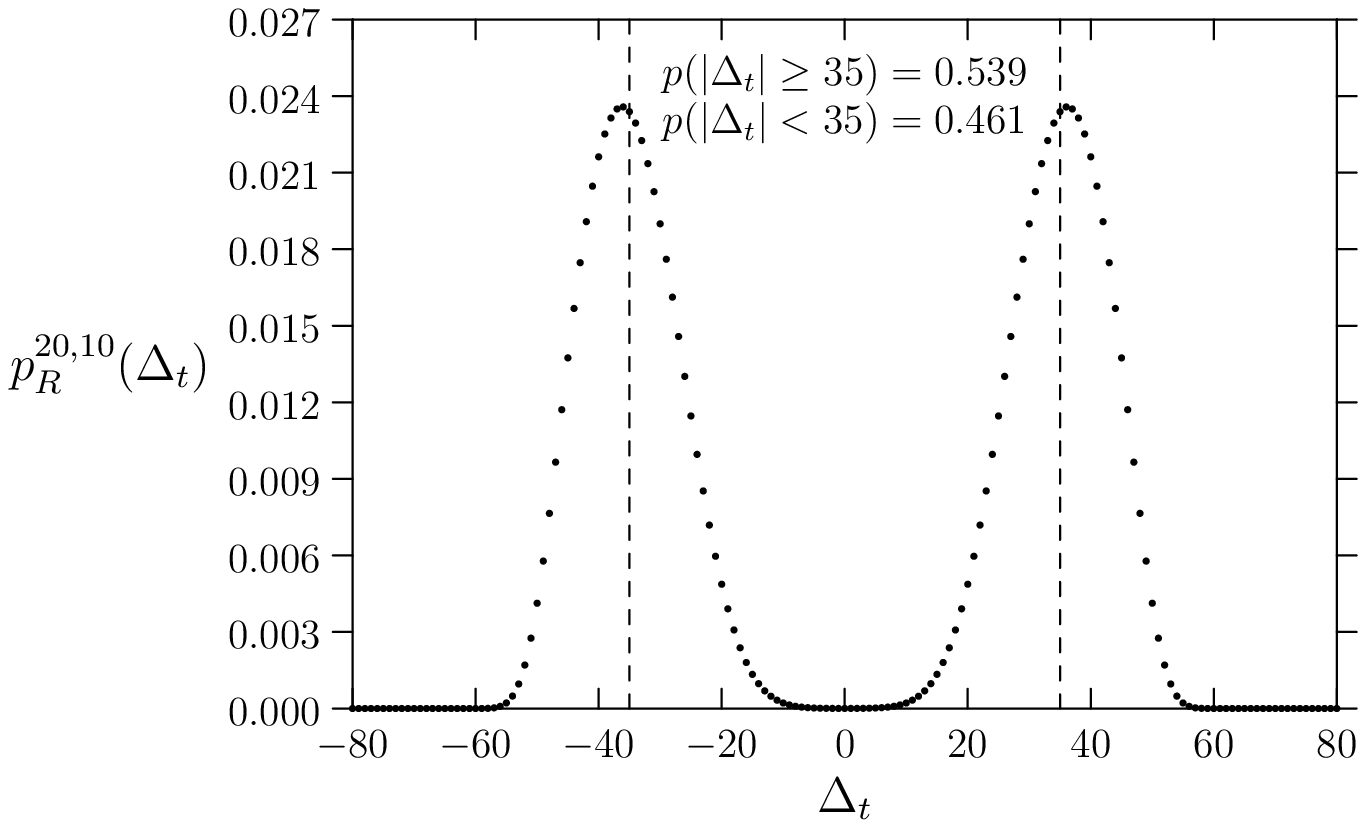}\\
\raisebox{3.5cm}{(c)}
\includegraphics[width=7cm]{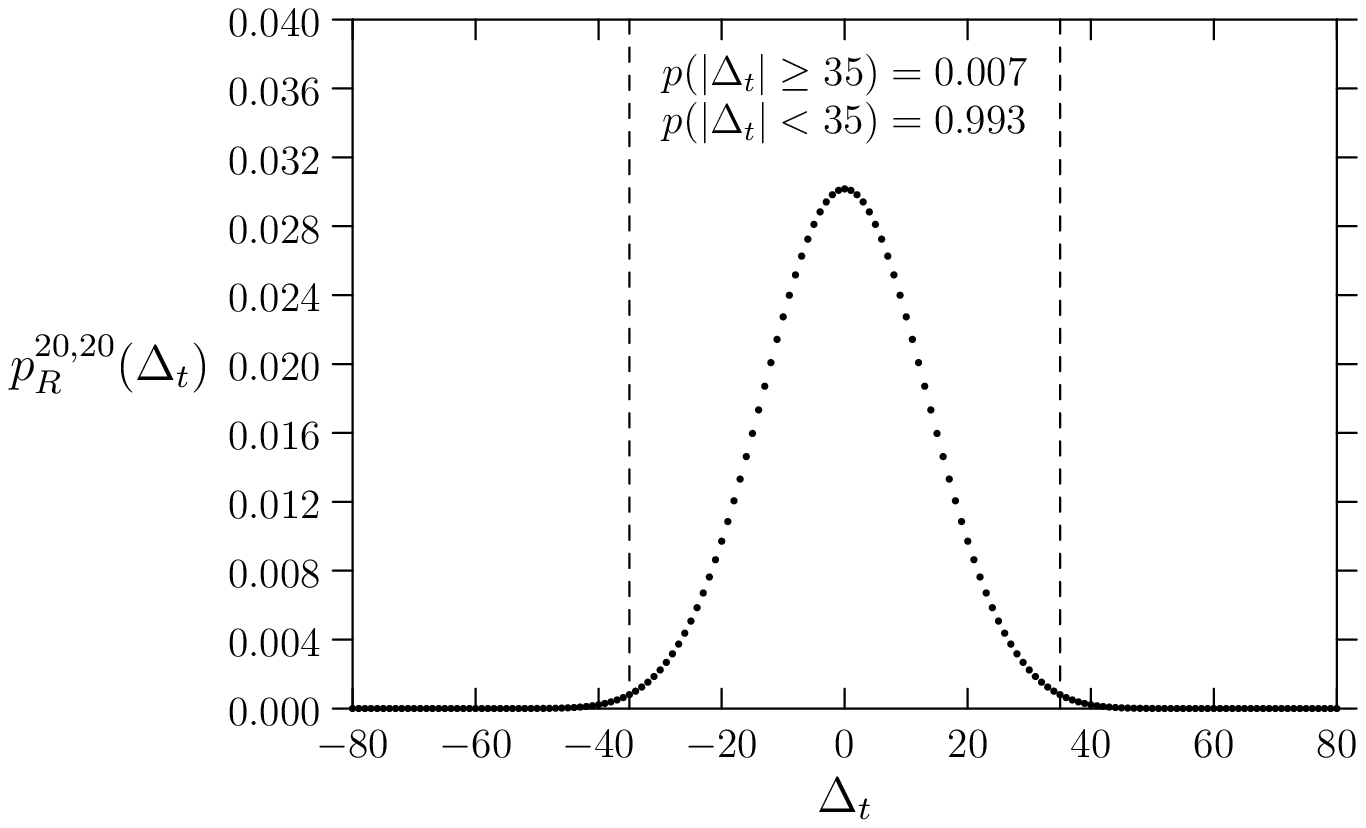}
\end{center}
\vskip-5mm
\caption{Distribution of the population difference $p^{S,\Delta}_R(\Delta_t)$ (Eq.~(\ref{plosses})) in the transmitted beam $t$ after the shutter for the state in Eq.~(\ref{example3}) subjected to $20\%$ of losses for $g=1.87$ assuming that $S=20$ photons were registered in the reflected beam and the difference measured by detectors was $\Delta=0$ (a), $\Delta=10$ (b), $\Delta=20$ (c).}
\label{ProbabilityDistributionsLosses}
\vskip-5mm
\end{figure}

\noindent
where $p^{S,\Delta=0}(k,l)$ is given by Eq.~(\ref{pkl}) and $\mathbf{S}$ is a set of $S$ for which the filter shutter is open, i.e.\ the probability of $|\Delta_t| \ge \delta_{th}$ evaluated for $\rho_{in}$ is greater than a given level of trust.  We chose $\delta_{th}=0$, $5$, $10$, $15$ and the level of trust $90\%$. The distribution $p_{\Phi}(k,l)$ and the corresponding distinguishabilities are depicted in Fig.~\ref{MacroPhotonNumber}. Although there is no clear separation between the regions $S_1$ and $S_2$ here, still, some low-probability gap appears which results in the increase of the distinguishability. For $\delta_{th}=0$, $5$, $10$ and $15$, the distinguishabilities are $0.72$, $0.93$, $0.96$ and $0.97$, respectively.

\subsubsection*{Multi-mode case and Losses}
\vskip-2mm

Let us consider two spatial or frequency modes in the input state in Eq.~(\ref{example3}). Since the two modes are independent, the probability distribution $p_2^{K,L}(k,l)$ resulting from detecting $K = n_1 + n_2$ and $L = m_1 + m_2$ photons in the detectors, where $n_1$ ($n_2$) and $m_1$ ($m_2$) are the contributions which come from the first (second) mode, is given by the convolution
\vskip-3mm
\begin{align}
&p_2^{K,L}(k,l) = \sum_{n_1 = 0}^K\sum_{m_1 = 0}^{L} \sum_{k_1=0}^k\sum_{l_1=0}^{l}\nonumber\\
&\qquad p^{n_1,m_1}(k_1,l_1)\,p^{K-n_1,L-m_1}(k-k_1,l-l_1).
\label{DistributionMode}
\end{align}

This distribution is depicted in Fig.~\ref{ProbabilityDistributionsMultiMode}. We note that the filtering process is deteriorated by the increase of the mode number. For the same parameters as in the single mode case ($g=1.87$, $S=K+L=20$, $\Delta=L-K=0$, $10$, $20$), but for lower threshold $\delta_{th}=35$ we achieved similar values of probabilities for a successful filtering $p(|\Delta_t| \ge 35)$ equal to: $0.553$, $0.459$, $0.061$ for $\Delta=0$, $10$, $20$, respectively.

Next, we computed the probability distribution $p^{S,\Delta}_R(\Delta_t)$ (Eq.~(\ref{plosses}) with $\tilde{\xi}_{nm}=\tilde{\gamma}_{nm}$ in Appendix D) for the state in Eq.~(\ref{example3}) subjected to $R=20\%$ of losses, see Fig.~\ref{ProbabilityDistributionsLosses}. Clearly, the filtering effect is preserved even for high losses. The higher gain and thus, the state population, the higher losses are tolerable. Effectively, losses diminish the available threshold values in comparison to the ideal case.

\section*{Appendix D: Losses}

The probability distribution $p^{S,\Delta}_R(\Delta_t)$ for the state in Eq.~(\ref{example3}) subjected to losses $R$ reads

\begin{align}
\nonumber\\
&p^{S,\Delta}_{R}(S_t,\Delta_t)=\sum_{n,m}\tilde{\xi}_{nm}\sum_{v=0}^n\sum_{w=0}^m f(v,w)
\nonumber\\&\quad
\sum_{n',m'}\tilde{\xi}_{n'm'}\sum_{v'=0}^{n'}\sum_{w'=0}^{m'} f(v',w')
\nonumber\\&\quad
\sum_{x=0}^{\min(n-v,n'-v')}\kern-2em{\tilde{c}}_x^{(n-v)}\,{\tilde{c}}_x^{(n'-v')} \delta_{n-v-x,n'-v'-x}  \delta_{n'-v'-x,\tfrac{S_t+\Delta_t}{2}}
\nonumber\\&\quad
\sum_{y=0}^{\min(m-w,m'-w')}\kern-2em{\tilde{c}}_y^{(m-w)}\,{\tilde{c}}_y^{(m'-w')}\delta_{m-v-y,m'-v'-y} \delta_{m'-v'-y,\tfrac{S_t-\Delta_t}{2}},
\label{plosses}
\end{align}
where
\begin{align}
f(v,w)=& \dfrac{c_v^{(n)}\,c_w^{(m)}}{\sqrt{v!\,w!\,2^{w+v}}}
\sum_{p=0}^v\sum_{q=0}^w\binom{v}{p}\binom{w}{q}(-1)^{v-p}
\nonumber\\&\quad
\delta_{\tfrac{S_r+\Delta_r}{2},v+w-\tfrac{S-\Delta}{2}}\,\delta_{\tfrac{S-\Delta}{2},p+q},\\
{\tilde{c}}_k^{(n)}=&\sqrt{\binom{n}{k}\,R^k\,(1-R)^{n-k}}.
\end{align}


\vfill




\begin{thebibliography}{99}

\bibitem{Sekatski2009} P. Sekatski, N. Brunner, C. Branciard, N. Gisin
  and C. Simon, Phys. Rev. Lett. {\bf 103}, 113601 (2009).

\bibitem{Sekatski2010} P. Sekatski, B. Sanguinetti, E. Pomarico,
  N. Gisin, and C. Simon, Phys. Rev. A {\bf 82}, 053814 (2010).

\bibitem{POVM} M. Nielsen and I. Chuang, {\it Quantum Computation and
    Quantum Information} (CUP, 2000).

\bibitem{Sanaka2006} K. Sanaka, K. J. Resch, and A. Zeilinger,
  Phys. Rev. Lett. {\bf 96}, 083601 (2006).

\bibitem{Resch2007} K. J. Resch, J. L. O'Brien, T. J. Weinhold,
  K. Sanaka, B. P. Lanyon, N. K. Langford and A. G. White,
  Phys. Rev. Lett. {\bf 98}, 203602 (2007).

\bibitem{DeMartini2008} F. De Martini, F. Sciarrino, and C. Vitelli, Phys. Rev. Lett. {\bf 100}, 253601 (2008).

\bibitem{Macrobell}T.~Sh.~Iskhakov, M.~V.~Chekhova, G.~O.~Rytikov, and G.~Leuchs, Phys. Rev. Lett. \textbf{106}, 113602 (2011).

\bibitem{Masha2} M.~Stobi\'nska, F.~T\"oppel, P.~Sekatski and M.~V.~Chekhova, Phys. Rev. A \textbf{86}, 022323 (2012).

\bibitem{Appel2008} J. Appel, E. Figueroa, D. Korystov, M. Lobino and
  A. I. Lvovsky, Phys. Rev. Lett. {\bf 100}, 093602 (2008).

\bibitem{Burks2009} S. Burks, J. Ortalo, A. Chiummo, X. Jia, F. Villa,
  A. Bramati, J. Laurat and E. Giacobino, Opt. Express {\bf 17}, 3777 (2009).

\bibitem{Gerasimov2011} L. V. Gerasimov, I. M. Sokolov, D. V. Kupriyanov, and M. D. Havey, arXiv:1111.6669.

\bibitem{Gisin2002} N. Gisin, G. Ribordy, W. Tittel and H. Zbinden,
  Rev. Mod. Phys. {\bf 74}, 145 (2002).
  
\bibitem{Vitelli2010-3} C. Vitelli, N. Spagnolo, L. Toffoli,
  F. Sciarrino, and F. De Martini, Phys. Rev. Lett. {\bf 105}, 113602
  (2010).

\bibitem{SpagnoloNoisy} N. Spagnolo, C. Vitelli, V. G. Lucivero, V. Giovannetti, L. Maccone, and F. Sciarrino, arXiv:1107.3726.

\bibitem{Vitelli2010-2} C. Vitelli, N. Spagnolo, F. Sciarrino, and
  F. De Martini, Phys. Rev. A {\bf 82}, 062319 (2010).

\bibitem{Sekatski2011} E. Pomarico, B. Sanguinetti, P. Sekatski,
  H. Zbinden, and N. Gisin, arxiv:1104.2212.

\bibitem{Stobinska09} M. Stobi\'nska, P. Horodecki, A. Buraczewski,
  R. W. Chhajlany, R. Horodecki, and G. Leuchs, arXiv:0909.1545.

\bibitem{Stobinska2011} M. Stobi\'nska, P. Sekatski, A. Buraczewski,
  N. Gisin, and G. Leuchs, Phys. Rev. A {\bf 84}, 034104 (2011).

\bibitem{Buraczewski2011} A. Buraczewski and M. Stobi\'nska, Comp. Phys. Commun. {\bf 183}, 2245 (2012).

\bibitem{Vitelli2010} C. Vitelli, N. Spagnolo, L. Toffoli,
  F. Sciarrino and F. De Martini, Phys. Rev. A {\bf 81}, 032123 (2010).

\bibitem{scissors} D. T. Pegg, L. S. Phillips, and S. M. Barnett,
  Phys. Rev. Lett. {\bf 81}, 1604 (1998).
  
\bibitem{DeMartini-PRL} F. De Martini, F.  Sciarrino, and Ch. Vitelli,
  Phys. Rev. Lett. {\bf 100}, 253601 (2008).

\bibitem{Masha} T. Iskhakov, M. V. Chekhova, and G. Leuchs,
  Phys. Rev. Lett. {\bf 102}, 183602 (2009).
  
\bibitem{Peres} A. Peres, {\it Quantum Theory: Concepts and Methods} (Kluwer Academic Publishers, Dordrecht, The Netherlands, 1993).

\bibitem{campos} R.A. Campos, B. E. A. Saleh and M. C. Teich, Phys. Rev. A {\bf 40}, 1371 (1989) 

%\bibitem{Kim2002} M. S. Kim, W. Son, V. Buzek, and P. L. Knight,
%  Phys. Rev. A {\bf 65}, 032323 (2002).

\end{thebibliography}
\end{document}